\documentclass[1p, 11pt, sort&compress,comma,square,numbers]{elsarticle}

\usepackage{graphicx}
\usepackage{wrapfig}
\usepackage{caption}
\usepackage{subcaption}
\usepackage{dcolumn}
\usepackage{bm}
\usepackage[mathlines]{lineno}
\usepackage[utf8]{inputenc}
\usepackage[T1]{fontenc}
\usepackage{newtxtext,newtxmath}
\usepackage{color}
\usepackage{setspace}
\usepackage{multirow}
\usepackage[normalem]{ulem}
\usepackage{algorithm}
\usepackage{algpseudocode}
\usepackage{titlesec}
\usepackage{enumitem}
\usepackage{siunitx} 
\usepackage{gensymb}
\usepackage{adjustbox}
\usepackage{booktabs}
\usepackage{makecell} 
\usepackage{fullpage}
\usepackage{hyperref}
\usepackage{hyperref}
\usepackage{makecell}
\hypersetup{colorlinks = true, linkcolor = blue, anchorcolor =red, citecolor = blue, filecolor = red, urlcolor = red, pdfauthor=author}

\usepackage[section]{placeins}
\begin{document}





\title{InSpecLearn4SDL: Interpretable Spectral Features Predict Conductivity in Self-Driving Doped Conjugated Polymer Labs}


\author[isuMechEAddress]{Ankush Kumar Mishra}
\author[NCSUAddress]{Jacob P. Mauthe}
\author[NCSUAddress]{Nicholas Luke}
\author[NCSUAddress]{Aram Amassian\corref{correspondingAuthor}}
\author[isuMechEAddress]{Baskar Ganapathysubramanian\corref{correspondingAuthor}}

\address[isuMechEAddress]{Department of Mechanical Engineering, Iowa State University, Ames, IA, 50010, USA}
\address[NCSUAddress]{Department of Materials Science and Engineering and ORaCEL, North Carolina State University,  Raleigh, NC, 27606, USA}
\cortext[correspondingAuthor]{Corresponding authors}
\date{\today}

\begin{abstract}

To accelerate materials discovery using self-driving labs (SDLs), we present a machine learning pipeline that predicts the electrical conductivity of doped conjugated polymers using rapid, non-destructive optical spectroscopy. Our approach automates spectral featurization by combining a genetic algorithm with adaptive area-under-the-curve (AUC) computations, creating a quantitative structure–property relationship (QSPR) that links optical response and processing parameters to conductivity. By incorporating SHAP-guided selection and domain-knowledge based feature expansion, the model matches expert-curated performance while theoretically reducing experimental effort by $\sim 33\%$ by minimizing the need for costly direct conductivity measurements. Notably, the model recovers known physical descriptors in pBTTT and identifies informative tail-state regions correlated with polymer bleaching upon successful doping. This generic, interpretable, small–data–friendly methodology can be potentially extended to other modalities, such as Raman or FTIR, providing a framework for autonomous decision-making in SDLs.

\end{abstract}

\maketitle

\section{Introduction}
\label{intro}

Conjugated polymers (CPs) have been investigated for a variety of organic electronics applications \cite{yu2020solution}, as well as emerging uses such as neuromorphic computing \cite{liu2021conjugated} and energy storage \cite{liang2015heavily}. CPs are organic macromolecules with backbones of alternating single and double bonds; the resulting delocalized $\pi$-electron cloud yields distinctive optical and electrical properties \cite{shirakawa1977synthesis, malik2023short, qiu2020conjugated}. As in inorganic semiconductors, doping is required to raise charge carrier density to useful levels \cite{kar2013doping, lu2021achieving}. The precise introduction of charge carriers has been central to advances in silicon technologies \cite{allen2019passivating, miao2012high} and, in organic electronics, is used to regulate charge transport for organic photovoltaics (OPVs) \cite{meerheim2014highly}, organic thermoelectrics (OTEs) \cite{bubnova2011optimization}, organic photodetectors \cite{siegmund2017organic}, organic light-emitting diodes (OLEDs) \cite{reineke2009white}, and organic field-effect transistors (OFETs) \cite{kim2023molecular, pei2022recent, lussem2013doped}.

Successful doping of CPs requires careful selection and synthesis of both the polymer and the dopant, and processing strongly influences physical state and properties \cite{kim2024disorder, kim2020highly}. Even within a single polymer–dopant system, numerous choices (solvents, annealing temperatures, doping times, environment) create a combinatorial design space. This combinatorial design space makes traditional experimentation resource-intensive, necessitating the use of laboratory automation and advanced statistical tools to navigate the diverse range of synthesis routes.

To systematically explore this space, scalable, automated synthesis and characterization are essential. Self-driving labs (SDLs) integrate optimization, machine learning (ML), and robotics to automate discovery \cite{rapp2024self, martin2023perspectives}. SDLs have been explored for thin-film properties \cite{liu2023autonomous, rooney2022self, macleod2020self, wang2025autonomous}, carbon nanotube synthesis \cite{nikolaev2016autonomy}, mechanics of additively manufactured objects \cite{gongora2021using, gongora2020bayesian}, nanoparticle synthesis \cite{zhao2023robotic, volk2023alphaflow,epps2020artificial}, yeast genetics \cite{king2004functional}, and catalyst composition \cite{burger2020mobile}, among other areas. SDLs address slow design-space exploration, gaps between experimental stages, and the absence of feedback to select subsequent experiments \cite{abolhasani2023rise}, using adaptive design of experiments (ADoE) to minimize experimental burden. They employ robotics for repetitive tasks and ML models as cost-effective surrogates for linking processing conditions to properties. Within SDLs, properties vary widely in evaluation cost. There is a strong interest in mapping inexpensive measurements to costly properties \cite{baishnab2024identifying}. Traditionally, surrogate features are identified by domain experts, yielding strong predictions but with system-specific, time-consuming efforts that do not readily generalize. As design complexity grows, reliance on manual intuition becomes a bottleneck.

A scalable alternative is to combine expert intuition with data-driven feature identification \cite{na2025interpretable}. Experts frame the physics and constraints; algorithms then explore broader candidate features, rank predictive power, and reveal non-obvious relationships. This hybrid approach leverages human insight and the speed and objectivity of ML, enabling more rapid, interpretable, and generalizable feature discovery.

For doped CPs, optical spectroscopy provides rich information before and after doping \cite{barford2017perspective}. Spectral signatures reflect phenomena such as polymer aggregation (linked to carrier mobility) \cite{boufflet2015using, wang2008solvent} and charge generation \cite{cochran2014molecular}. Conductivity obeys $\sigma = |e| \mu n$, where $\sigma$ is electrical conductivity, $|e|$ is the elementary charge magnitude, $\mu$ the mobility, and $n$ the carrier concentration. Spectroscopy is fast (seconds to a minute) and non-destructive, preserving samples for further processing. Thus, spectral features are attractive surrogates for building quantitative structure–property relationships (QSPRs) linking structure and processing to conductivity. QSPRs have been applied across domains \cite{hu2010review, fayet2016use, kwon2019comprehensive, fluetsch2024adapting, tinkov2020cross, liu2020understanding, wang2023sustainable, wen2021simultaneous}.

While raw, pointwise spectra are ideal in principle \cite{zhang2023editors}, they are often impractical in low-data regimes due to their high dimensionality. Spectral featurization is a viable alternative. For X-ray absorption near-edge spectra (XANES), prior work has used cumulative distribution function (CDF), peak-based descriptors, and wavelet transforms with dimensionality reduction (PCA, Isomap, autoencoders) \cite{timoshenko2009wavelet, munoz2003continuous, chen2024robust, manthiram2020reflection}. For UV–Vis, raw absorbance with PCA/PLS has been employed \cite{razavi2023ultraviolet, gariso2025comparative}. Latent representations via autoencoders have been explored for spectrum–structure relationships in catalysts \cite{routh2021latent}. Torrisi \textit{et al.} \cite{torrisi2020random} improved interpretability by transforming X-ray absorption spectra into multiscale polynomial features that capture local trends. Yoon \textit{et al.} \cite{yoon2024explainable} used B-splines-based descriptors to featurize the UV-vis-NIR spectra and used a coefficient shrinkage regression model, LASSO, to identify important regions of the UV-vis-NIR spectra for conductivity prediction of doped conjugated polymers.

Each method has trade-offs: raw spectra are unwieldy at small dataset sizes; peak features can be sensitive to noise; and dimensionality reduction methods may lose information, typically benefiting from larger datasets. We address these challenges with a featurization strategy based on the area under the curve (AUC) combined with a genetic algorithm (GA). AUC over adaptively selected windows encodes feature magnitude and width while being more noise-robust; GA identifies informative regions for downstream modeling.

\begin{figure*}
	\centering
\includegraphics[width=1\linewidth]{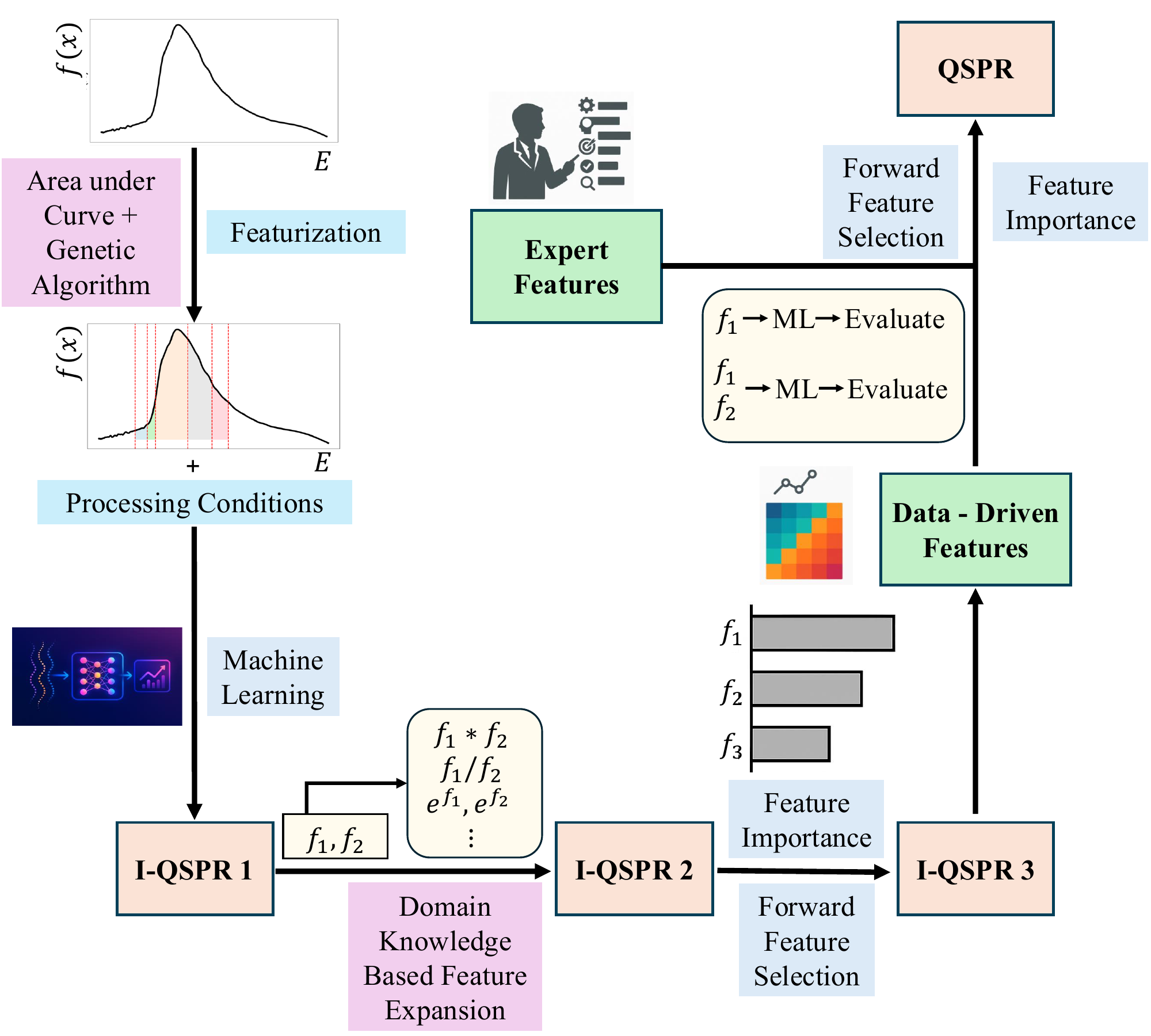}
    \caption{Workflow for generating a QSPR model that maps optical spectra and processing conditions to electrical conductivity. Spectral features are extracted using the area under the curve (AUC), and key regions are identified using a genetic algorithm. These features are used to train the initial model, QSPR 1. To enhance performance, mathematical operations are applied to expand the feature set, resulting in QSPR 2. Feature importance is then assessed, and greedy forward selection is employed to identify a compact, high-performing subset, termed data-driven features, yielding QSPR 3. Expert-curated features are subsequently incorporated to develop the final QSPR model. In the absence of expert input, QSPR 3 serves as the final model. The data-driven features are also interpreted and benchmarked against expert-selected features.}
    \label{Fig:graph_abstract}
\end{figure*}

We treat the derived features as surrogates for conductivity and build a QSPR via data-driven feature engineering, benchmarking against a baseline with expert-curated features. The data-driven model matches the expert-guided model, and a hybrid (data-driven + expert) model outperforms both, highlighting the value of integrating human intuition with ML. Our methodology is generic and can identify informative regions in optical spectra and, more broadly, can be potentially applied to other spectral modalities (XANES, Raman, FTIR). These regions can then be used to predict a quantity of interest (QoI), provided the spectra are physically representative of that QoI.

\textbf{Our contributions:} Our key contributions to this work include the following:
\begin{itemize}\itemsep0em
    \item \textbf{Data-driven spectral featurization}: We propose a data-driven method to featurize optical spectra using the AUC with optimization (GA), and develop a QSPR model for predicting conductivity in doped conjugated polymers. 
    \item \textbf{Feature engineering}: We perform feature engineering to identify key, interpretable features and demonstrate that the data-driven model achieves predictive performance comparable to models based on expert-identified features. 
    \item \textbf{Human machine learning collaboration}: We combine data-driven and expert features to develop a hybrid model that outperformed individual models, demonstrating the benefit of integration human intuition with machine learning.
    \item \textbf{Theoretical reduction in experimental time}: We show that conductivity characterization accounts for a measured 33\% of the total experimental time. By using optical spectra as inputs, these labor-intensive steps can be theoretically eliminated, potentially enabling a 33\% reduction in the total experimental cycle time.

\end{itemize}

\section{Results and Discussion}
\label{sec:results}

\subsection{Data Collection}

\subsubsection{Processing Conditions}

For this study, we focus on a well-known model system, pBTTT as the conjugated polymer and F4TCNQ as the dopant administered through the dip-doping process. The primary reason for choosing this system is the well-established spectral analysis \cite{wang2008solvent, yamashita2019efficient}, which will be used as a baseline for comparison later in the study. Using the materials chosen, we first need to constrain the formulation and processing variables to a reasonable number of experimental conditions by identifying suitable cosolvents for pBTTT using the computed Hansen solubility parameters (HSP). We selected a subset of solvents based on prior literature showing that the choice of solvent strongly influences aggregation and thereby the carrier mobility of pBTTT-based organic field-effect transistors (OFETs) \cite{wang2008solvent, yi2017solvent}. We selected three solvents, namely chlorobenzene (CB), ortho-dichlorobenzene (DCB), and toluene (Tol), as these showed more than an order of magnitude variation in field-effect mobility \cite{wang2008solvent, yi2017solvent}. We further constrained the processing parameter space using differential scanning calorimetry (DSC) data and established crystallization dynamics of pBTTT \cite{pirela2024crystallization} to determine the optimal window of annealing temperatures, between room temperature and 270 °C. This range encompasses multiple phase transitions and yields morphologically diverse films when combined with the mixing of the aforementioned solvents. While other parameters, such as dip-doping solvent and annealing temperature of the doped film, could influence performance, our study focused on varying the cosolvent composition of the pBTTT solution and the annealing temperature of the resulting film. Accordingly, the processing conditions considered in this work are the percentages of CB, DCB, and Tol, as well as the annealing temperature. Several other processing conditions were held fixed to focus on the role of polymer processing and its effect on polymer microstructure. These include the polymer concentration (5 mg/mL), spin coating conditions (1500 rpm),  doping solvent of n-Butyl Acetate (nBA), the concentration of F4TCNQ in this solution (2 mg/mL), and a post-doping annealing temperature (60\(\degree\)C).


\subsubsection{Experimental Setup}

\begin{figure*}[t!]
	\centering
\includegraphics[width=1\linewidth]{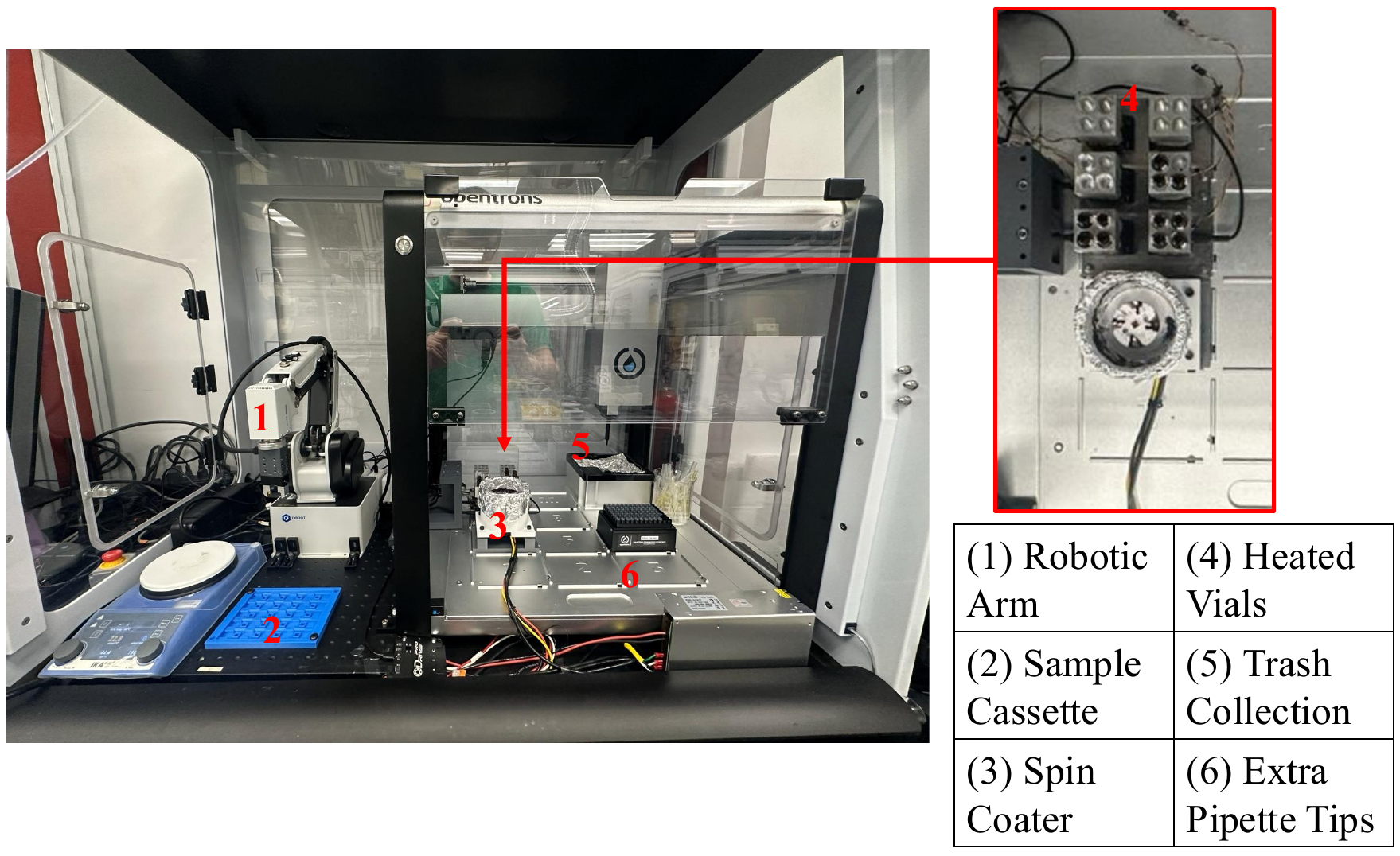}
    \caption{Materials acceleration platform (MAP) used for preparation of polymer films, highlighting the robotic sample manipulation, multi-sample cassette, computer-controlled spin coater, and heated vial storage.}
    \label{Fig:exp_setup}
\end{figure*}

 The experimental platform used for processing the films is shown in Fig. \ref{Fig:exp_setup}. The platform is a Materials Acceleration Platform (MAP), developed at North Carolina State University. It is comprised of an Opentrons OT-2 pipetting robot, a computer-controlled spin coater with a custom 3D-printed housing designed to fit into the Opentrons, and modified MHP30 mini hot plates used for solution heating. A Dobot MG400 robotic arm is used for substrate and sample manipulation. The mini hotplates were outfitted with custom-machined aluminum blocks, which enabled the heating of four vials per hotplate, a necessity for high-temperature spin coating, “hot casting”. Hot casting is a requirement for solution-processing pBTTT, which has been shown to otherwise gel at room temperature \cite{patel2017morphology,yi2019pbttt}. While the MAP is not yet fully self-driving, several steps in the experimental workflow are already automated.

\begin{figure*}[t!]
	\centering
\includegraphics[width=1\linewidth]{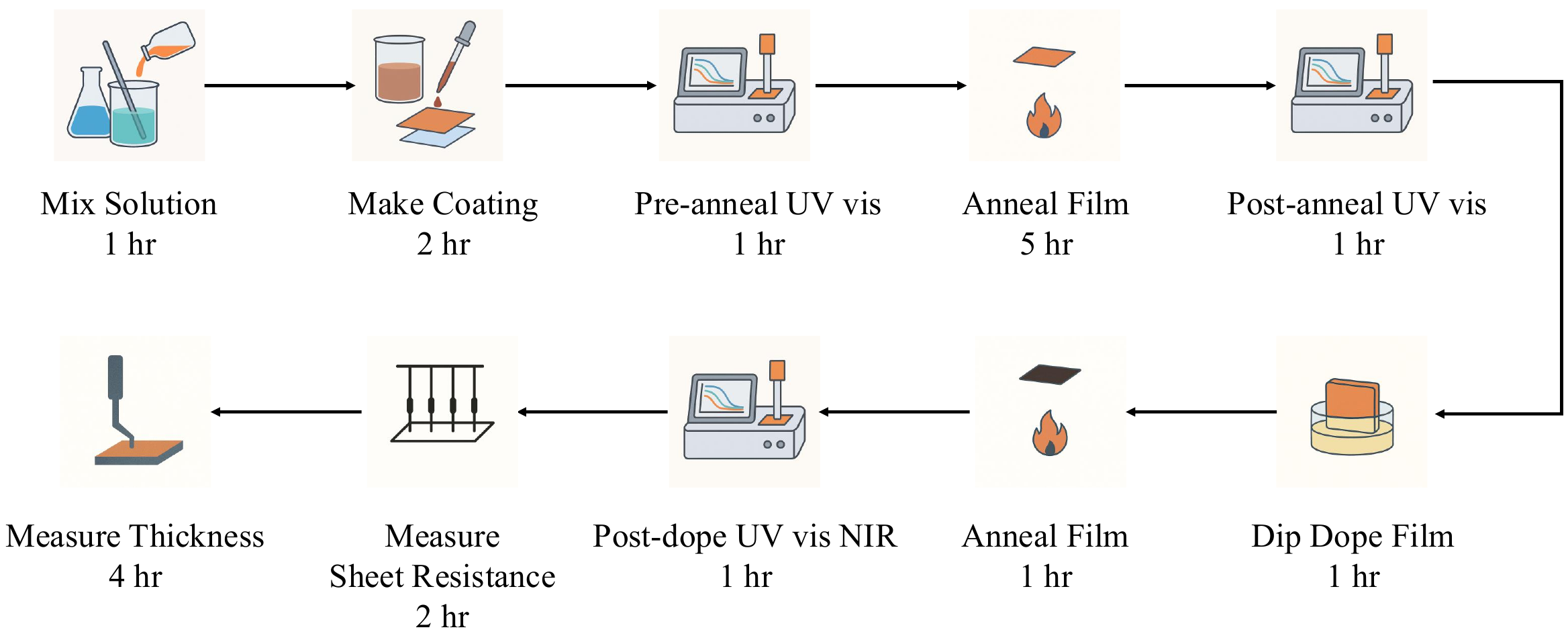}
    \caption{Workflow for processing, doping, and characterizing a batch of doped conjugated polymer films. The steps include solution preparation, film coating, sequential spectroscopic measurements, annealing, doping, and final conductivity characterization. The timeline for each step is shown for a batch of 32 samples, highlighting that conductivity measurements are the most time-consuming stage.}
    \label{Fig:time_analysis}
\end{figure*}

Figure~\ref{Fig:time_analysis} illustrates the step-by-step workflow for preparing a set of 32 samples with duplicates, collecting the spectroscopy, and measuring their conductivity. The process begins with the automated mixing of pBTTT precursor solutions to give the desired co-solvent mixture using the Opentrons platform, followed by automated spin coating. Optical spectroscopy is then performed on the as-cast films, after which the samples undergo annealing. Following annealing, another round of optical spectroscopy is conducted to capture any changes in the spectroscopic signatures that may have occurred during annealing. The film is then doped using a dip-doping method and annealed again. A final spectroscopy step is performed on the doped films. Lastly, sheet resistance and thickness measurements are carried out, which are used to calculate conductivity. Three measurements were taken from both duplicate samples and averaged for statistical robustness.

We perform the experiments on 128 samples. The 128 samples are selected using Bayesian Optimization (BO) for efficient exploration of the design space.  We start with 32 samples, obtained through Latin Hypercube sampling (LHS), and fit a Gaussian process regression (GPR) between the processing conditions and conductivity. We then use the Upper Confidence Bound acquisition function to select the next batch of 32 samples. We perform 3 batches of BO to obtain a total of 128 samples (32 from LHS and 96 from BO). Further details about the BO process, collection, and sharing of data between multi-disciplinary laboratories can be found in our other papers ~\cite{Mauthe2026AI, tali2025sears}. 

Figure~\ref{Fig:time_analysis} reports the time required to process a batch of 32 samples at each step. Conductivity measurement (comprised of the sheet resistance and thickness measurements) accounts for one-third of the total experimental duration. Specifically, measuring thickness via stylus profilometry is destructive and labor-intensive, requiring manual scraping and multiple readings per sample. Successfully predicting conductivity from optical signatures could eliminate these two operational steps, theoretically reducing experimental effort by $\sim33\%$ and substantially increasing the throughput of automated experimentation.

\subsection{Data Partitioning: Train Test Split}
Our dataset consists of 128 samples, obtained through Bayesian exploration of the design space, each corresponding to a unique combination of processing conditions and their corresponding electrical conductivity. A common approach to splitting data is to perform a random data split between the train, validation, and test sets. However, for smaller datasets, this can lead to uneven distributions between the train, validation, and test sets, resulting in biased evaluation. 

\begin{figure*}[t!]
    \centering

    \begin{subfigure}[c]{0.35\linewidth}
        \centering
        \adjustbox{valign=m}{
            \includegraphics[width=\linewidth]{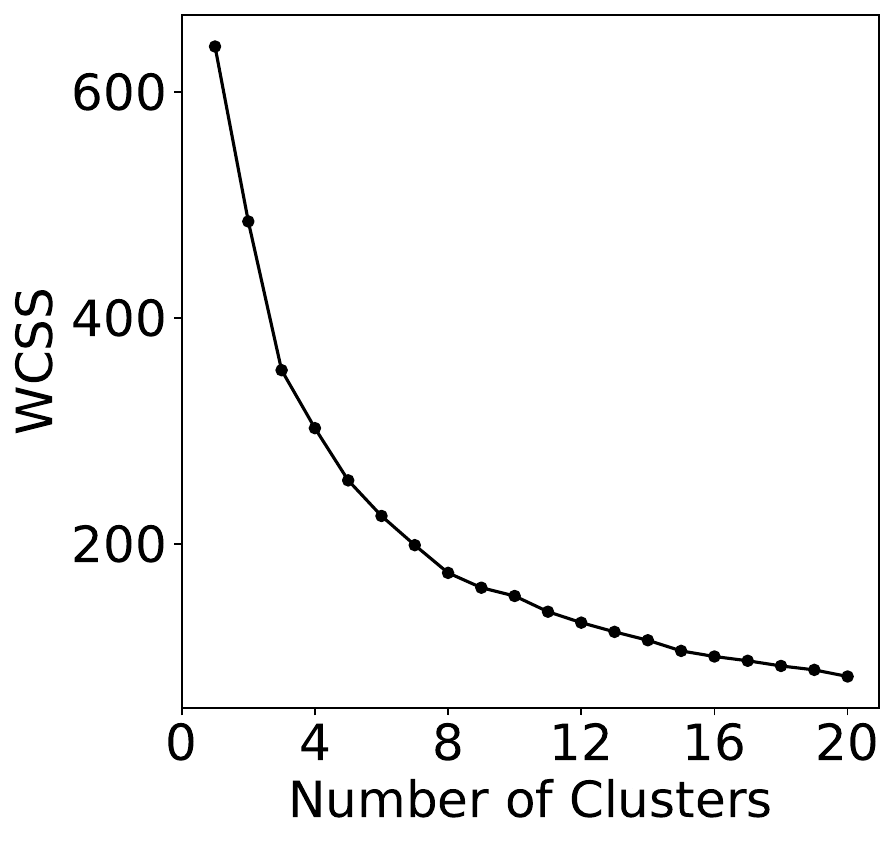}
        }
        \caption{}
        \label{Fig:elbow}
    \end{subfigure}
    \hfill

    \begin{subfigure}[t]{\textwidth}
        \centering
        \includegraphics[width=\linewidth]{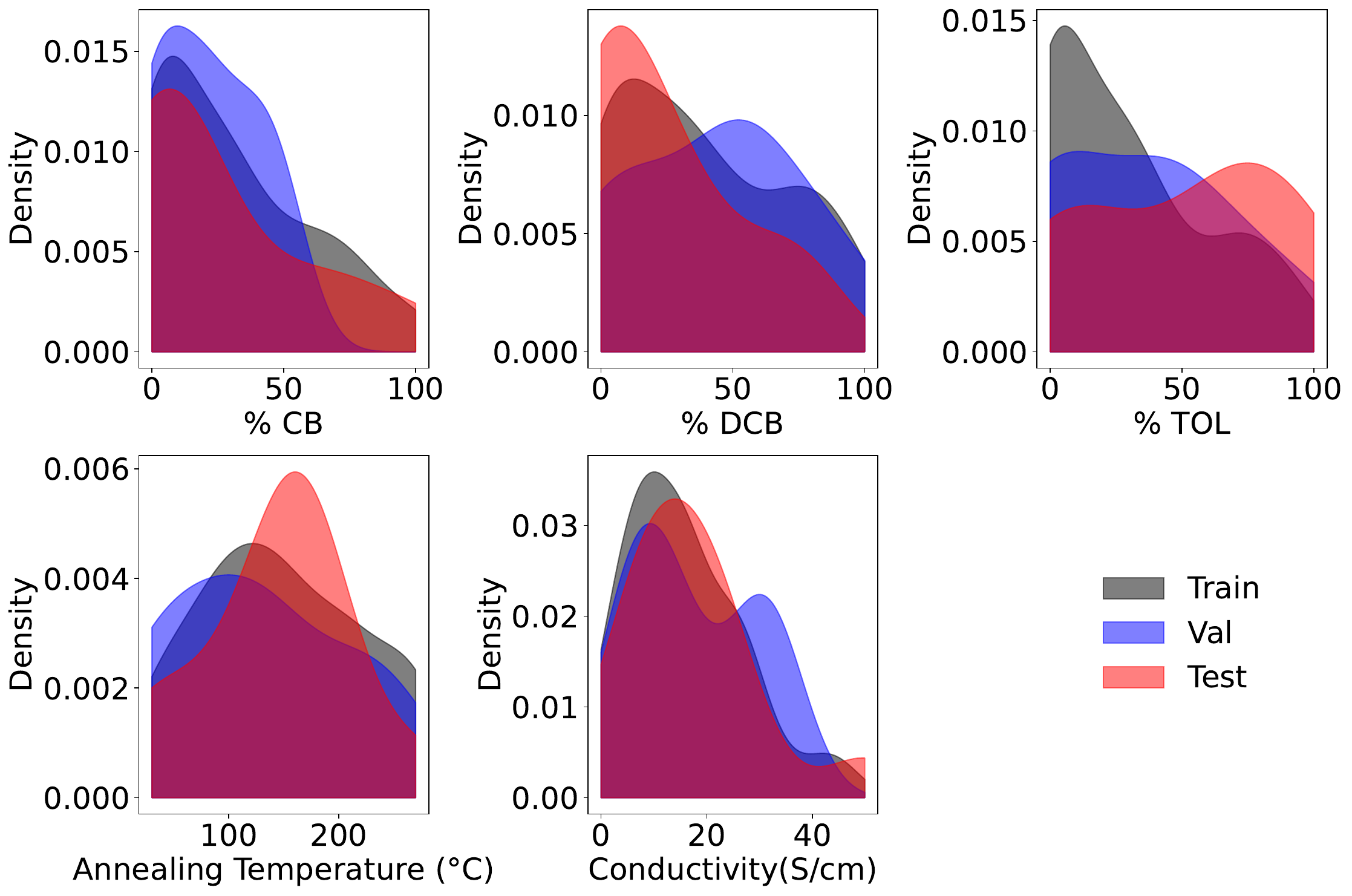}
        \caption{}
        \label{Fig:kde}
    \end{subfigure}

    \caption{Data distribution analysis using clustering, KS test, and KDE plots. (a) Elbow method for selecting the optimal number of clusters. The plot displays the within-cluster sum of squares (WCSS) against the number of clusters. The "elbow" point, where the rate of decrease in WCSS slows down, indicates the optimal number of clusters  (b) Kernel Density Estimation (KDE) plots comparing the distributions of processing conditions and conductivity training and test datasets. }
    \label{Fig:data_distribution_analysis}
\end{figure*}

To avoid this, we first cluster the data to capture its structure. We utilize K-means clustering and determine the optimal number of clusters using the elbow method. The elbow method utilizes the within-cluster sum
of squares (WCSS) distance to identify the optimum number of clusters. It does so by finding the "elbow point", which corresponds to the number of clusters that slows down the decrease in WCSS distance. The optimum number of clusters identified using the elbow method was 5, as shown in Figure~\ref {Fig:elbow}. From each cluster, we randomly selected 20\% of the data points, corresponding to 5 points per cluster. These 25 data points are then randomly divided into two sets: a validation set and a test set. The remaining 103 points form the training dataset. The test dataset is kept separate to prevent any data leakage in subsequent model training. 

To confirm that all three sets follow the same distribution, we use the Kolmogorov–Smirnov (KS) test ~\cite{ks-test} which compares their empirical distributions. The KS test evaluates the following hypotheses:

\begin{equation}
\begin{aligned}
    \text{Null Hypothesis } (H_0): & \quad F(x) = G(x) \\
    \text{Alternative Hypothesis } (H_A): & \quad F(x) \neq G(x)
\end{aligned}
\end{equation}
where \( F(x) \) and \( G(x) \) represent the distribution of the training and test datasets, respectively.

From Table~\ref{tab:ks} (Appendix~\ref {appendix2}), we observe that all p-values are greater than the significance threshold of $\alpha = 0.05$. Hence, we fail to reject the null hypothesis $H_o$, indicating that the training and test data are drawn from the same distribution. This supports the assumption that the training, validation, and test data sets should originate from the same underlying data distribution, which is central to most ML models.

\subsection{Featurization of Spectra and Identification of Optimum Bin Locations}
\label{sec:featurize_results}

To utilize the spectral data, we need to extract meaningful features from the raw spectra collected during the experimental process. These spectra represent three different physical states of the film: as-cast (or unannealed), post-annealed, and post-dope. The as-cast spectra will provide insight into the effects of co-solvent mixtures. As previously noted, the processing solvent may influence aggregation of the polymer film, resulting in noticeable changes to the polymer's absorbance spectrum, such as vibronic progressions. The post-annealed spectra will therefore be more informative about the effects that annealing has on further aggregating (or deaggregating) the polymer as a function of temperature. We expect that this will be more informative than the as-cast spectra due to the strong influence of thermal history and crystallization dynamics. Finally, we expect the post-dope spectroscopy to be informative about the doping process itself. Here we can look for differences in polymer bleaching, anion spectra, and polaron spectra that may be indicative of fluctuations in carrier concentration, which could impact the conductivity \cite{kwon2024quantifying,kiefer2019double}. We also preprocess the raw spectra by performing min-max normalization followed by curve smoothing using the Savitzky-Golay filter function from \textit{SciPy}. The raw spectra cannot be directly used for model training due to the limited dataset size, and features based on peak or valley position, width, and intensity are highly sensitive to noise. This makes it challenging for algorithms to reliably distinguish true spectral features from noise-induced artifacts.

\begin{figure*}[t!]
	\centering
\includegraphics[width=1\linewidth]{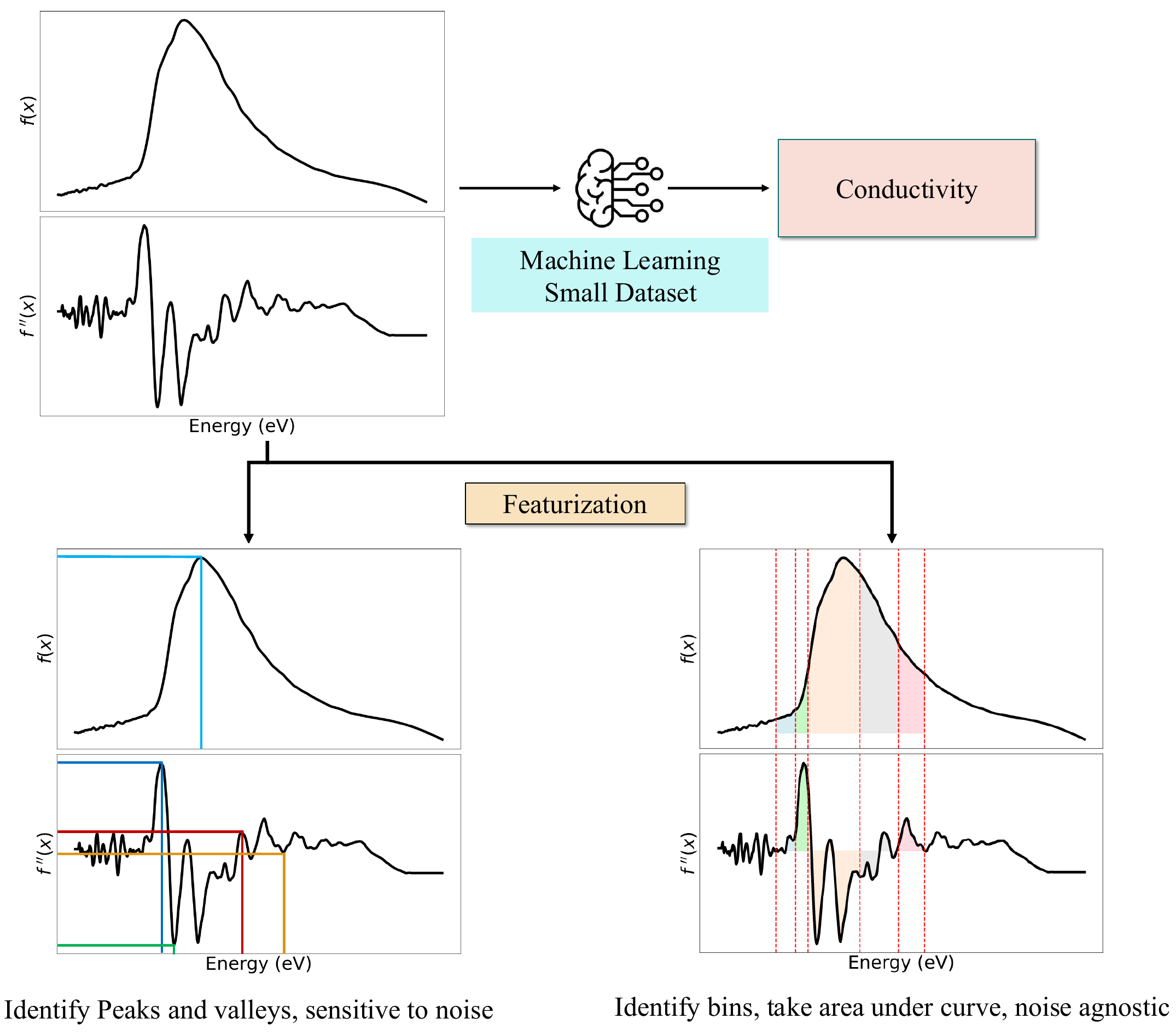}
    \caption{Featurization of optical spectra for conductivity prediction in doped conjugated polymers. Peak and valley-based features are sensitive to noise, whereas binning followed by calculating the area under the curve offers a more noise-robust approach. }
    \label{Fig:featurization_type}
\end{figure*}

As discussed in Section \ref{intro}, AUC can serve as a robust alternative feature. Figure \ref{Fig:featurization_type} shows how we can use AUC features as an alternative to the identification of peaks and valleys. The AUC captures both the magnitude and the spread of spectral features, implicitly accounting for peak (and valley) intensity, width, and position, while being less sensitive to noise compared to discrete peak/valley detection. To apply this method, we divide the spectrum into a set of bins (identified by the bin locations), and the AUC within each bin is computed as a feature. 

The choice of bin locations is critical; well-placed bins isolate informative spectral regions and suppress noisy or irrelevant segments. We cast bin selection as a black-box optimization problem over ordered bin boundaries. This objective is non-convex and non-differentiable; the area-under-the-curve (AUC) features change discretely as boundaries cross peaks/shoulders, and the fitness depends on downstream model training and cross-validation, making gradient-based methods ill-suited.

\begin{figure*}[t!]
	\centering
\includegraphics[width=1\linewidth]{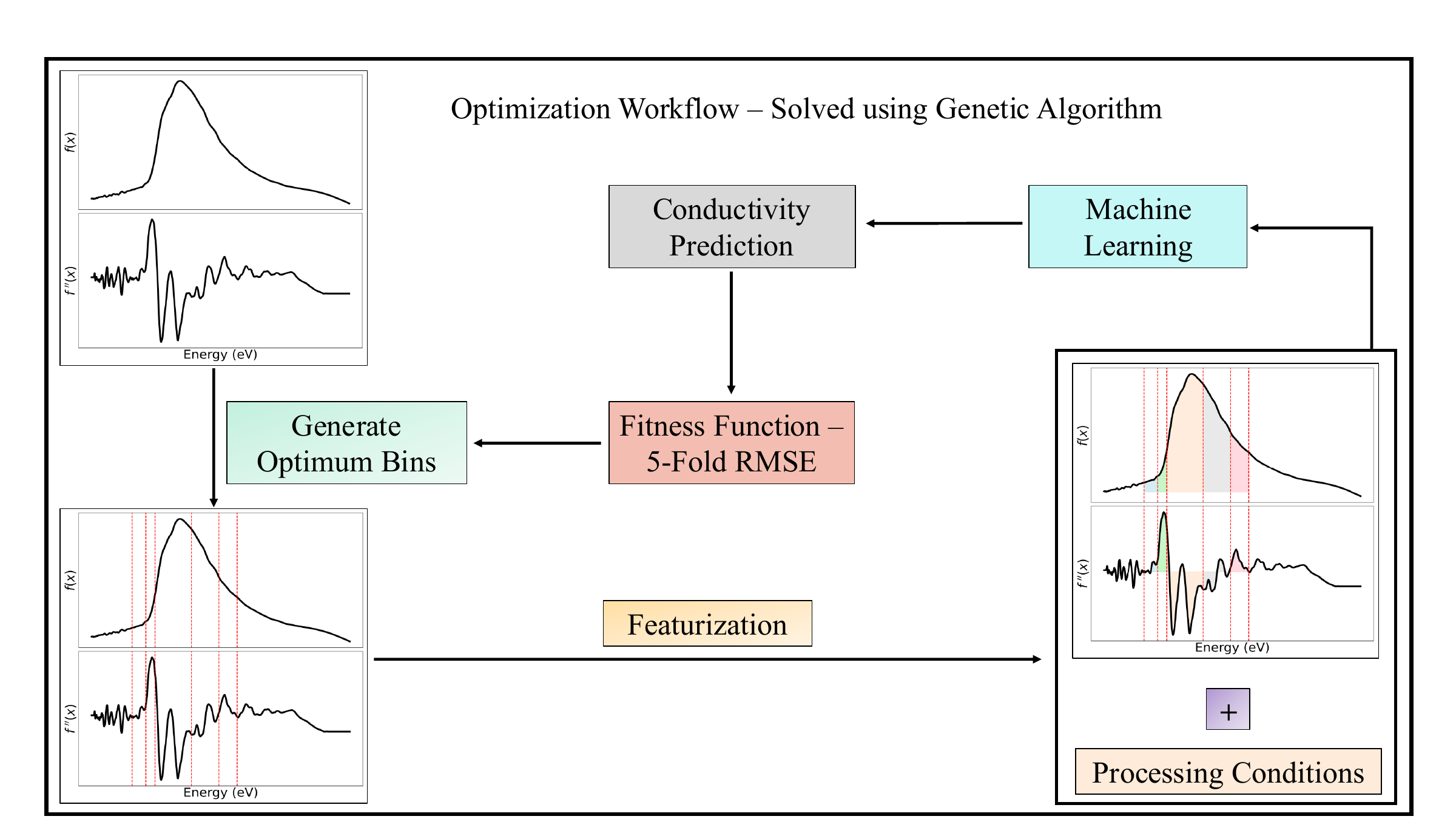}
    \caption{Workflow for Genetic Algorithm-based spectral bin optimization. Processing conditions and optical spectra are used to generate features through a GA-driven binning strategy. The GA optimizes bin locations by minimizing the 5-fold cross-validated RMSE of a machine learning model trained to predict conductivity. The resulting features are then used for training the final model and predicting conductivity.}
    \label{Fig:ga_setup}
\end{figure*}

We therefore use a genetic algorithm (GA) to identify an optimal set of bin locations (see workflow in Fig. \ref{Fig:ga_setup}). We use the training dataset solely to identify the optimal bin locations, thereby avoiding data leakage. GA is a population-based, derivative-free global search method inspired by the principles of natural selection. Rather than following local gradients, it maintains a diverse population of candidate solutions and uses selection, crossover, and mutation to explore the search space across generations. This makes GA less prone to getting trapped in a single local minimum than single-start, gradient-driven optimizers. In our encoding, each candidate represents an ordered set of bin boundaries constrained to lie within the spectral domain; ordering is essential because AUC is computed between consecutive boundaries. We also enforce a minimum bin width to avoid degenerate intervals. The fitness of a candidate is the cross-validated predictive score obtained when AUC features from its bins (optionally combined with processing parameters) are used to train the model.

Several hyperparameters govern GA behavior. The population size controls how broadly the space is explored; the crossover probability encourages exploitation by recombining high-fitness candidates; the mutation probability injects diversity to probe new regions; and the number of generations sets the search horizon (with diminishing returns after a point). We use a population of 100, a crossover probability of 0.7, a mutation probability of 0.3, and 100 generations, following common heuristics and prior practice \cite{greenstein2023determining}. We repeat the GA multiple times with different seeds. While the exact bin locations varied, the selected spectral regions for featurization were consistently similar.

The fitness of each solution, analogous to a loss function, is evaluated through the following
process:
\begin{itemize}
    \item For each optical spectrum, we compute the AUC under each bin of the candidate. 
    \item We then compute the AUC for the second derivative of the spectra. The choice of the second derivative, in addition to the original spectrum, was based on domain knowledge. The second derivative is calculated from the min-max normalized raw spectra. We then use the Savitzky-Golay filter function from \textit{SciPy} and set the "deriv" parameter to 2.
    \item Then we combine the AUC features from the original and second derivative spectra with the corresponding processing parameters. As a guiding principle, we aim to keep the total number of features for the ML model to roughly 10-15\% of the training dataset size to avoid overfitting. As the training dataset size was 103, we experimented with 4, 5, and 6 bin locations—corresponding to 3, 4, and 5 bins respectively—yielding 6, 8, and 10 AUC features (from both the original and second-derivative spectra). Among these, the best model performance was observed using 5 bin locations. However, the results and the important features identified for 4 and 6 bin locations were qualitatively similar, suggesting stability in feature selection across a reasonable range of bin counts.
    \item After this, we train an ML regression model using the training dataset to predict conductivity. We chose a random forest regression model. A detailed discussion of the choice of regression model is presented in Section \ref{sec:qspr1}.
    \item Finally, we evaluate the model by computing 5-fold cross-validation root mean square error (RMSE) between predicted and true conductivity for the training dataset. RMSE is used as the fitness function to be minimized.
\end{itemize}

In each generation of GA, the creation of the population proceeds as below-

\begin{itemize}
    \item The top $p\%$ of the current population (elite solutions) are passed unchanged to the next generation to preserve high-performing candidates. We set $p = 5\%$.
    \item $q\%$ of the new population is generated using crossover and mutation. We set $q = 45\%$:
    \begin{itemize}
        \item Tournament selection is used to choose parents for crossover and mutation. This is done by selecting multiple random candidates from the current population and choosing among them based on their fitness value. This ensures randomness while also ensuring that we choose the best parent among the random candidates.
        \item Crossover involves swapping portions of bin locations between two parents at a randomly selected crossover point. The resulting offspring are sorted to maintain the constraint that the bin locations in a candidate should be in increasing order. 
        \item Mutation perturbs one or more bin locations within a solution by a random value in a user-defined range.
    \end{itemize}
    \item The remaining $(100-p-q)\%$ (or 50\%) of the population is filled with newly generated random candidates to encourage exploration.
    
\end{itemize}

\subsubsection{Analysis of Spectra and Interpretation of Optimum Bin Locations}

 Through the featurization of the three different spectra for all samples, we identify that the most informative features consistently come from the post-anneal spectra. There are likely several factors that lead to the pre-anneal (as-cast) and post-dope spectra providing less predictive power, including the processing parameters chosen and the physical changes that happen during doping. In the case of the former, we observe that the annealing temperature serves as the single most influential processing parameter. While the pre-anneal spectra will reflect sample-to-sample differences due to the co-solvent mixture, the thermal history of the sample from the annealing step has a dominating effect, causing much of the information stored in the pre-annealed spectra to lose significance after the annealing has been performed. This naturally leads to the post-anneal spectrum, which contains the most pertinent information about polymer structure and aggregation prior to doping, emphasizing both the role and predictive power of the pseudo-"structural analysis" that featurization provides. On the other hand, post-doping spectra could be expected to be the most informative with regard to conductivity predictions because they are taken while the sample is in the same physical state as the conductivity measurements. Although it is true that the post-dope spectra contain the most information about the doping process itself (such as carrier concentration), they also lose valuable information about the polymer structure and order due to the bleaching that occurs during the doping process. The ground-state electrons responsible for the absorption of the undoped polymer are transferred to the dopant during the doping process, and thus, any physical insight they could provide also disperses. Due to the fixed dip-doping conditions of 2 mg/mL dopant in nBA for 10 minutes, there is much less sample-to-sample variation to observe in the post-doping spectrum. Due to the significantly higher predictive power of the post-anneal spectra, we shift our focus to features from that spectrum going forward.

\begin{figure*}[ht]
    \centering

    \begin{subfigure}[b]{0.48\textwidth}
        \includegraphics[width=\textwidth]{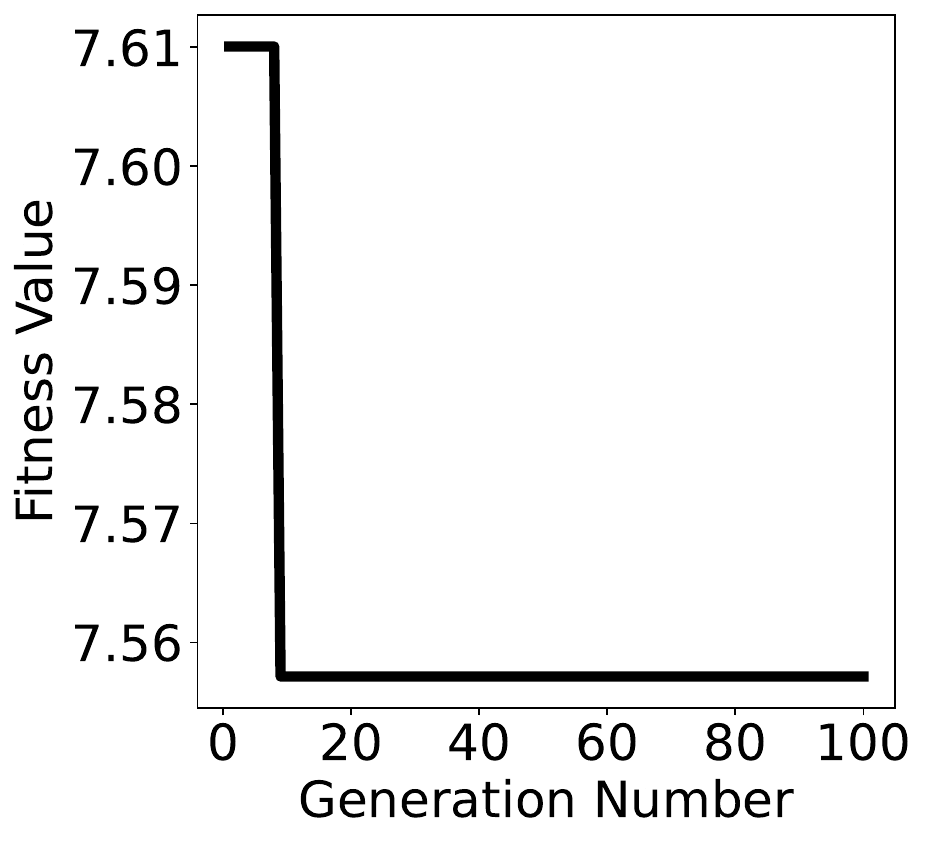}
        \caption{}
        \label{fig:ga_training}
    \end{subfigure}
    \hfill
    \begin{subfigure}[b]{0.48\textwidth}
        \includegraphics[width=\textwidth]{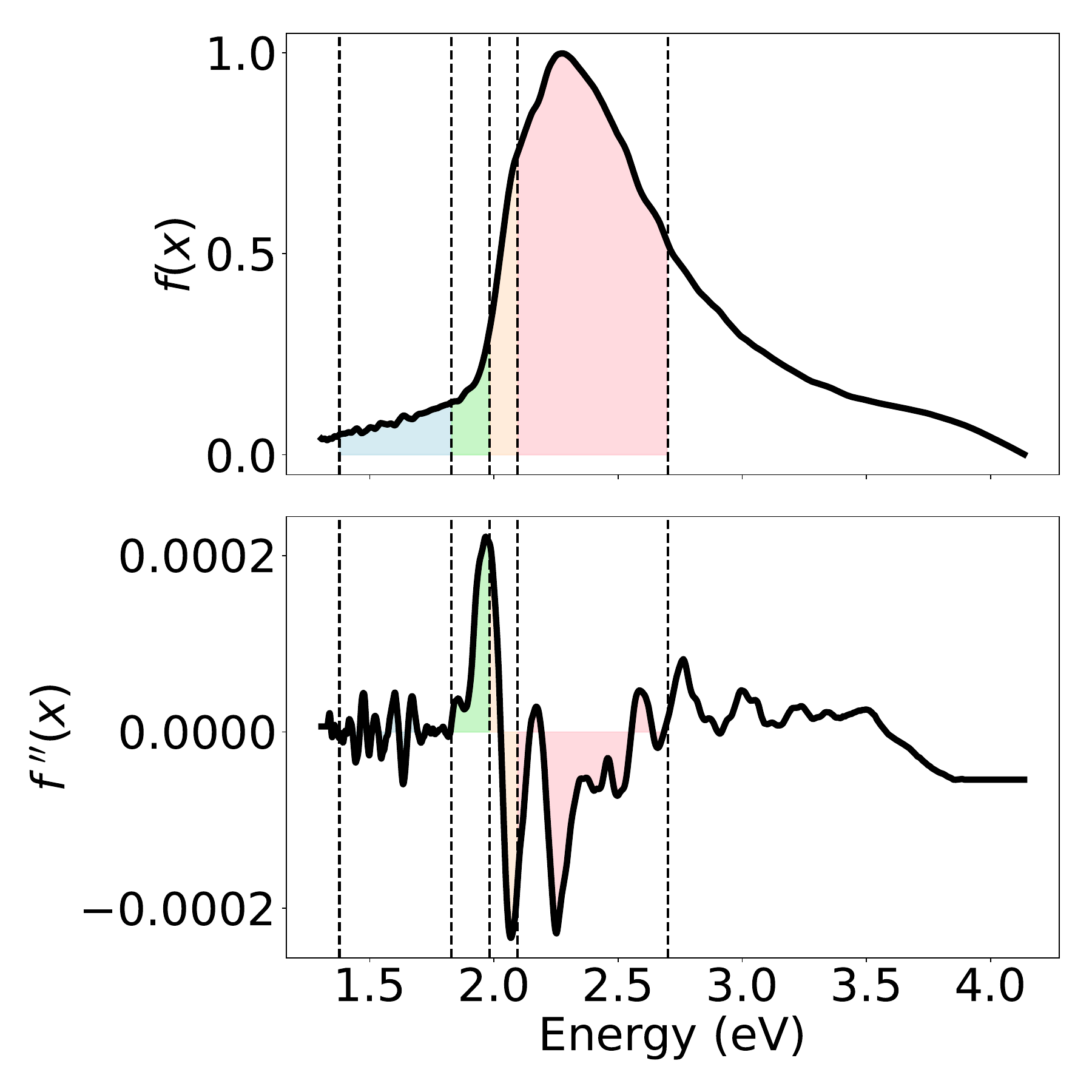}
        \caption{}
        \label{fig:optimum_bins}
    \end{subfigure}

    \caption{(a) Fitness value progression over generations during genetic algorithm optimization.  (b) Optimal bin locations identified by the genetic algorithm, overlaid on the absorbance spectrum (top) and its second derivative (bottom). Shaded regions represent the spectral segments selected for AUC feature extraction, and vertical red lines denote the bin boundaries.}
    \label{fig:ga_results}
\end{figure*}

Figure \ref{fig:ga_training} shows the fitness value across the 100 generations using GA. The optimal bin locations in the post-anneal spectra identified by GA were [1.378, 1.828, 1.982, 2.095, 2.700] eV as shown in Figure \ref{fig:optimum_bins}.
These bins represent energy intervals where meaningful spectral changes occur, correlating with conductivity. These bin locations contain meaningful information about the polymer's aggregation when analyzed in the right context. The low-energy bin, from 1.378-1.828 eV, lies in the sub-gap region of the absorbance spectrum and thus reflects the tail states arising from the polymer's semi-crystalline nature.
The second bin, from 1.828 to 1.982 eV, contains the onset of the 0-0 vibronic peak. The AUC of this bin in the original spectrum and its second derivative will contain some information about the shifting of the peak position, reflecting potential red- or blue-shifting. 
The third bin, from 1.982-2.095 eV,  actually contains the 0-0 vibronic transition, which corresponds to an electronic excitation without a change in the molecular vibrational state. The varying of this feature's prominence in the second derivative AUC will reflect red-shifting or blue-shifting of this low-energy transition and indicate differences in the ground-state energy, likely arising from variations in aggregation or structural order. Similarly, the AUC from the original spectrum will reflect the relative prominence of the 0-0 transition compared to other spectral features, which should correspond to the well-studied 0-0/0-1 ratio.  
The final bin, from 2.095-2.700 eV, contains the high-energy 0-1 and 0-2 vibronic transitions. The AUC from this region will contain information relevant to the 0-0/0-1 ratio, and the second derivative will reflect the positioning of these transition energies.  

Combining all of these bins together, a detailed profile of the polymer's excited state emerges: the 0-0 transition reveals information about the ground state, the 0-1 transition elucidates the strength of electron-vibration coupling, and information from the 0-2 transition would allow for quantification of these interactions through calculation of optoelectronic parameters \cite{wang2008solvent}. Further, the ratio of various features, for example, the 0-0/0-1 ratio, has been previously shown to indicate exciton delocalization and the degree of solid-state ordering, which are relevant for doped carrier mobility  \cite{zhao2013entanglement}. A physical explanation for each of the terms used in this paragraph has been provided in Appendix \ref{appendix6}.

\subsection{Intermediate QSPR Model 1}
\label{sec:qspr1}

Once the optimal bin locations (candidate) are identified using GA, we compute the AUC for both the optical spectra and their second derivatives using these bins. Table \ref{tab:feature_descriptions} lists all 8 features and their description. These spectral features are then combined with the corresponding processing conditions to form the complete input feature set. Using this feature set, we train a variety of regression models and evaluate their performance. We explored several categories of algorithms: linear algorithms (Linear Regression, LASSO, Ridge), tree-based ensemble algorithms (Random Forest and Gradient Boosting), as well as Support Vector Regression, K-Nearest Neighbors, and Gaussian regression. Among these, tree-based ensemble algorithms consistently provided the best predictive performance. Table \ref{tab:all_model_results} (Appendix \ref{appendix2}) shows the performance of various algorithms.

Tree-based models outperformed linear alternatives by effectively capturing the nonlinear interactions and feature couplings inherent in doped conjugated polymer systems. Unlike linear models, which often require extensive feature engineering to handle complex dependencies, tree-based methods automatically learn hierarchical decision rules across categorical and continuous data. This approach is particularly advantageous in our workflow as it requires minimal preprocessing and remains robust to outliers, a critical factor given that conductivity can vary by two orders of magnitude due to processing variations.

To assess how well the model generalizes to unseen samples, we use a combination of evaluation metrics: $R^2$, RMSE, Mean Absolute Error (MAE), Kendall Tau correlation, and Pearson correlation. Each metric provides insight into different aspects of model performance in the context of predicting electrical conductivity. $R^2$ quantifies how well the model explains the variance in measured conductivity compared to a simple baseline that always predicts the mean conductivity. RMSE emphasizes larger errors, making it relevant for identifying whether the model fails on outlier samples, such as those samples with unusually high or low conductivity. MAE provides the average magnitude of prediction error, offering a more robust and interpretable measure of accuracy across the dataset, regardless of outliers. Kendall Tau correlation measures the agreement in ranking between predicted and true conductivity values. Pearson correlation captures the strength of the linear relationship between predicted and actual conductivity values. Together, these metrics provide a comprehensive evaluation, capturing how much variance the model explains, its sensitivity to extreme cases, and how well it preserves both the direction and scale of conductivity trends.

\begin{figure}[ht]
    \centering
    \begin{subfigure}[t]{0.19\textwidth}
        \centering
        \includegraphics[width=\linewidth]{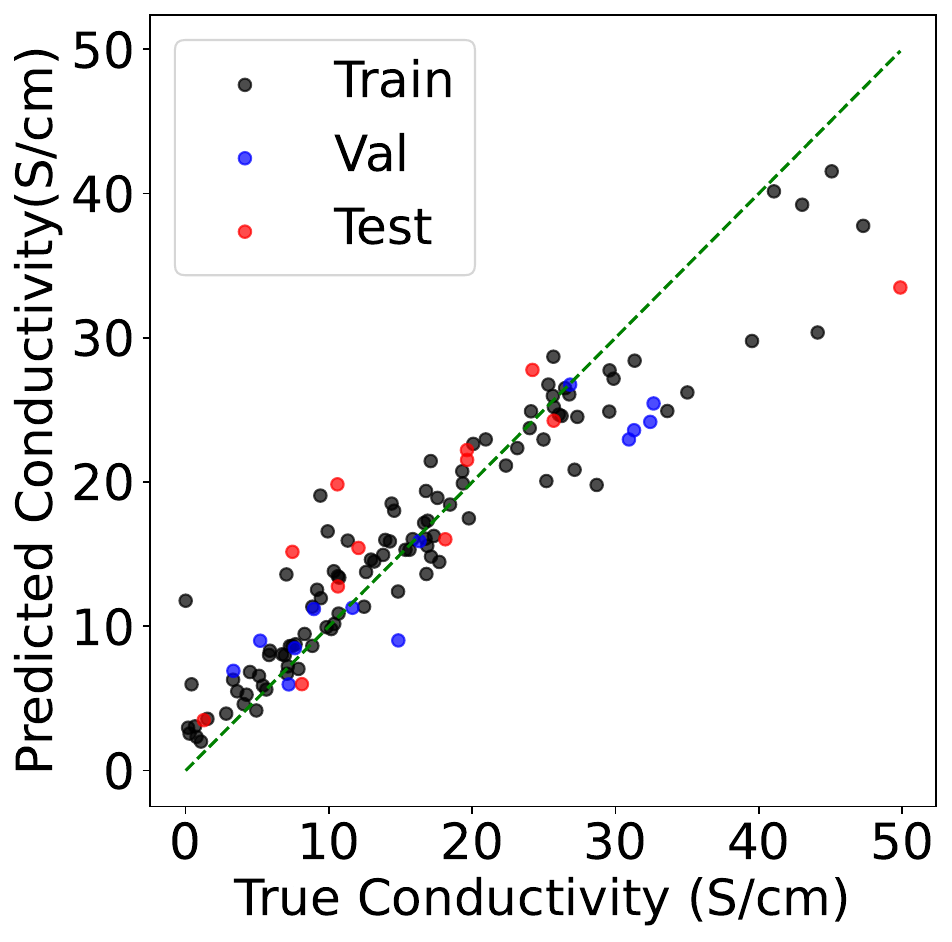}
        \caption{I-QSPR 1}
        \label{fig:model1_plot}
    \end{subfigure}
    \hfill
    \begin{subfigure}[t]{0.19\textwidth}
        \centering
        \includegraphics[width=\linewidth]{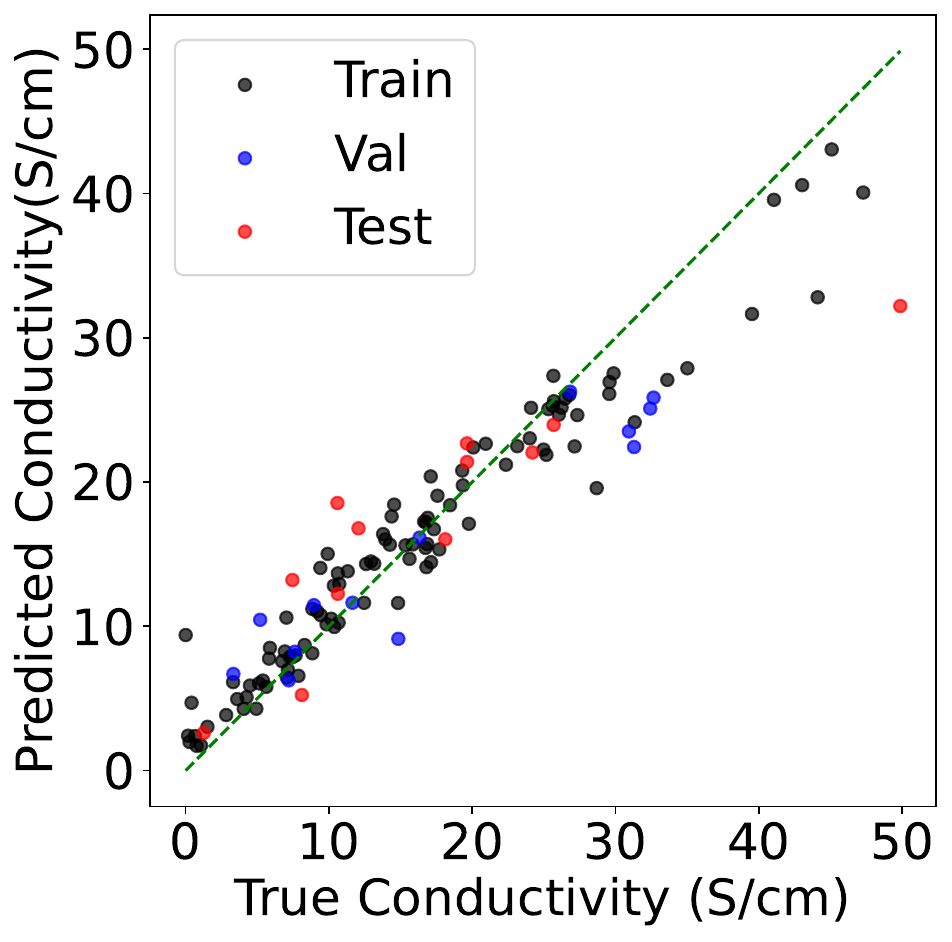}
        \caption{I-QSPR 2}
        \label{fig:model2_plot}
    \end{subfigure}
    \hfill
    \begin{subfigure}[t]{0.19\textwidth}
        \centering
        \includegraphics[width=\linewidth]{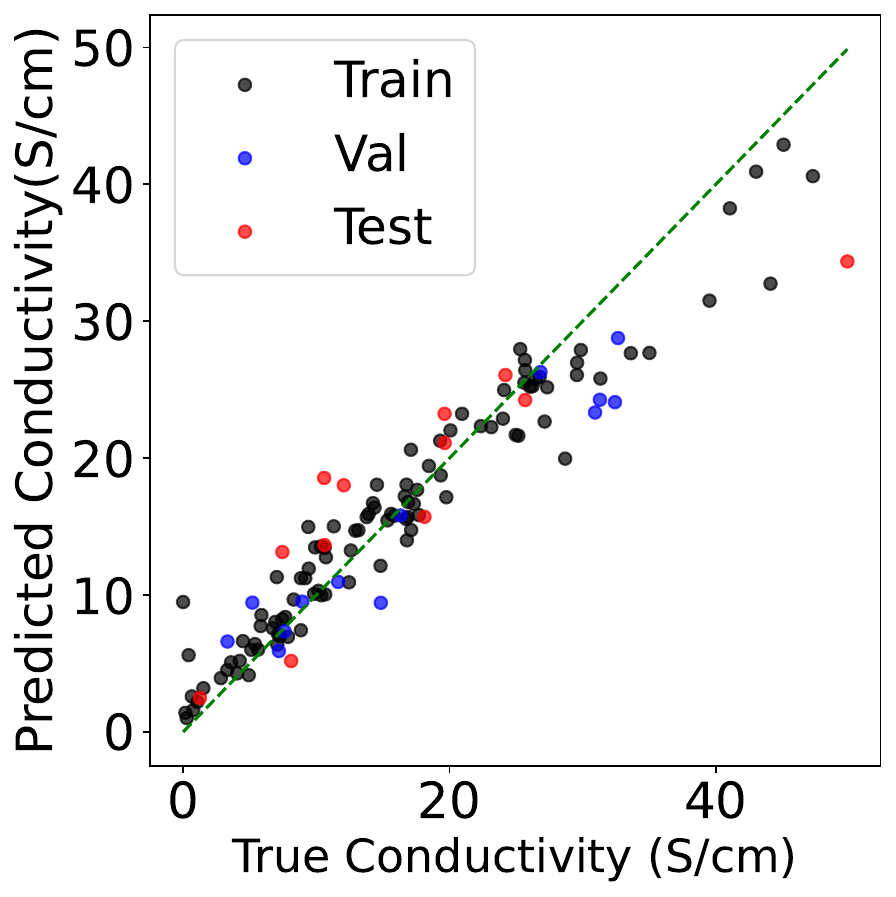}
        \caption{I-QSPR 3}
        \label{fig:model3_plot}
    \end{subfigure}
    \hfill
    \begin{subfigure}[t]{0.19\textwidth}
        \centering
        \includegraphics[width=\linewidth]{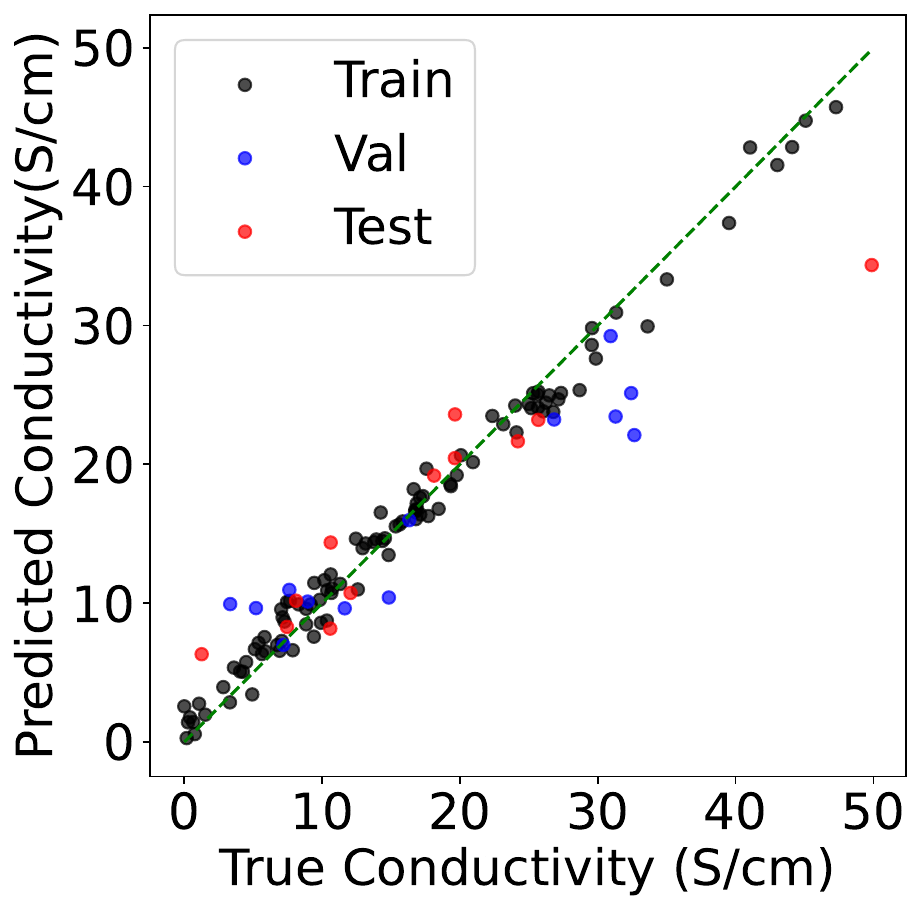}
        \caption{E-QSPR}
        \label{fig:model4_plot}
    \end{subfigure}
    \hfill
    \begin{subfigure}[t]{0.19\textwidth}
        \centering
        \includegraphics[width=\linewidth]{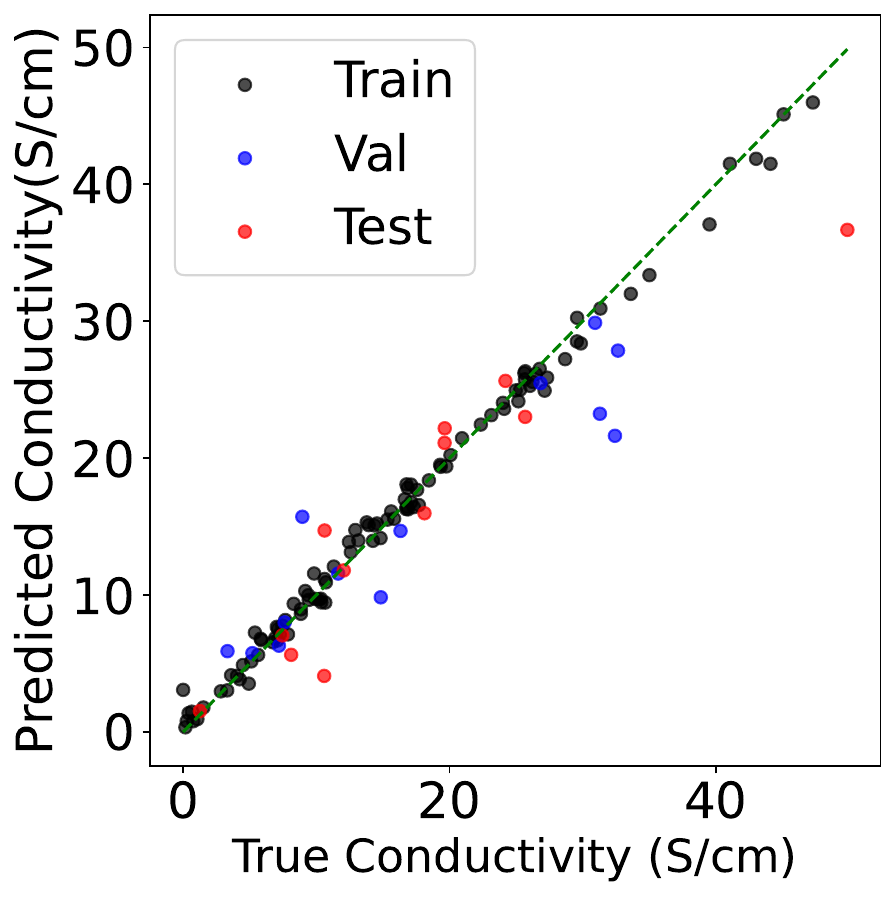}
        \caption{QSPR}
        \label{fig:model5_plot}
    \end{subfigure}
    \vspace{1em}
    \caption{ QSPR Models: Combined regression results and evaluation metrics. (a - e) True conductivity vs predicted conductivity for train and test dataset using I-QSPR Model 1, 2, 3, E-QSPR, and final QSPR}
    \label{fig:model1_combined}
\end{figure}

We evaluated various algorithms for intermediate QSPR 1 (Table \ref{tab:all_model_results}). Among them, the Random Forest model yielded the best predictive performance. Figure~\ref{fig:model1_plot} shows the predicted versus true conductivity values for both the training, validation, and test sets. The performance metrics for the QSPR models are summarized in Table~\ref{tab:model1_results}. On the test set, the model achieved an $R^2$ score of 73.17\%, indicating strong generalization and confirming the predictive capability of features derived from adaptively binned optical spectra.

\begin{table}[ht]
    \centering
    \caption{Abbreviations and descriptions of processing conditions, spectral AUC features, derivative AUC features, and product terms used in this study}
    \label{tab:feature_descriptions}
    \begin{tabular}{cc}
        \Xhline{1.2pt}
        Feature & Description \\
        \Xhline{1.2pt}
        CB & \% of Chlorobenzene solvent (processing condition)\\
        DCB & \% of Ortho-dicholorobenzene solvent (processing condition)\\
        Tol & \% of Toulene solvent (processing condition)\\
        annealing\_temperature & Annealing temperature (\degree C) of as-cast film (processing condition)\\
        \Xhline{0.8 pt}
        AUC\_1 & AUC of original spectra between 1.378-1.828 eV \\
        AUC\_2 & AUC of original spectra between 1.828–1.982 eV \\
        AUC\_3 & AUC of original spectra between 1.982-2.095 eV \\
        AUC\_4 & AUC of original spectra between 2.095-2.700 eV \\
        \Xhline{0.8pt}
        $d^{2}\mathrm{AUC}\_{1}$ & AUC of second derivative of spectra between 1.378-1.828 eV \\
        $d^{2}\mathrm{AUC}\_{2}$ & AUC of second derivative of spectra between 1.828-1.982 eV \\
        $d^{2}\mathrm{AUC}\_{3}$ & AUC of second derivative of spectra between 1.982-2.095 eV \\
        $d^{2}\mathrm{AUC}\_{4}$ & AUC of second derivative of spectra between 2.095-2.700 eV \\
        \Xhline{0.8pt}
        X*Y & Product between feature X and Y. X and Y can be any of the 8 AUC features above\\ 
        \Xhline{1.2pt}
    \end{tabular}
\end{table}

\begin{table}[ht]
    \centering
    \caption{{QSPR Models' Performance Metrics for Training, Validation and Test Set}}
    \label{tab:model1_results}  

    \resizebox{\textwidth}{!}{%
    \begin{tabular}{ccccccccccc}
        \Xhline{1.2pt}
        \multirow{2}{*}{Type} & \multirow{2}{*}{Model} & \multirow{2}{*}{Data} & \multirow{2}{*}{Algorithm} & \multirow{2}{*}{Input} & \multirow{2}{*}{Output} & $R^2$  & RMSE & MAE & Kendall  & Pearson  \\
        & & & & & & (\% $\uparrow$) & ($\downarrow$)& ($\downarrow$) & Tau (\% $\uparrow$) & (\% $\uparrow$) \\
        \Xhline{1.2pt}
        \multirow{9}{*}{\shortstack{Data\\Driven}} 
        & \multirow{3}{*}{I-QSPR 1} & Train & \multirow{3}{*}{\makecell{Random\\Forest}} & \multirow{3}{*}{AUC,$p$} & \multirow{3}{*}{$\sigma$} 
            & 88.84 & 3.68 & 2.52 & 83.31 & 95.49 \\
        & & Val & & & & 80.28 & 4.90 & 3.81 & 76.92 & 95.50 \\
        & & Test  &  & & & 73.17 & 6.25 & 4.56 & 78.79 & 88.20\\
        \Xcline{2-11}{0.8pt}
        & \multirow{3}{*}{I-QSPR 2} & Train & \multirow{3}{*}{\makecell{Random\\Forest}} & \multirow{3}{*}{AUC,$p$, $M$} & \multirow{3}{*}{$\sigma$} 
            & 92.55 & 3.00 & 2.12 & 86.32 & 97.23 \\
            & & Val & & & & 80.20 & 4.91 & 3.81 & 71.79 & 95.24 \\
        & & Test  &  & & & 73.18 & 6.25 & 4.39 & 75.76 & 88.74\\
        \Xcline{2-11}{0.8pt}
        & \multirow{3}{*}{I-QSPR 3} & Train & \multirow{3}{*}{\makecell{Random\\Forest}} & \multirow{3}{*}{$D$} & \multirow{3}{*}{$\sigma$} 
            & 92.68 & 2.98 & 2.12 & 85.79 & 96.99 \\

        & & Val & & & & 84.02 & 4.41 & 3.36 & 74.36 & 96.17 \\
        & & Test  &  & & & 76.09 & 5.90 & 4.42 & 78.79 & 89.52 \\
        \Xhline{1 pt}
        \multirow{3}{*}{Expert} 
        & \multirow{3}{*}{E-QSPR } & Train & \multirow{3}{*}{\makecell{Gradient\\Boosting}} & \multirow{3}{*}{$E$} & \multirow{3}{*}{$\sigma$} 
            & 98.39 & 1.40 & 1.14 & 92.07 & 99.37 \\
            & & Val & & & & 78.40 & 5.13 & 4.12 & 61.54 & 94.15 \\
        & & Test  &  & & & 81.49 & 5.19 & 3.49 & \textbf{84.85} & \textbf{94.53}  \\
        \Xhline{1 pt}
        \multirow{3}{*}{Combined} 
        & \multirow{3}{*}{QSPR} & Train & \multirow{3}{*}{\makecell{Gradient\\Boosting}} & \multirow{3}{*}{$C$} & \multirow{3}{*}{$\sigma$} 
            & 99.31 & 0.91 & 0.68 & 94.32 & 99.72 \\
            & & Val & & & & 81.07 & 4.80 & 3.32 & 71.79 & 92.79 \\
        & & Test  &  & & & \textbf{85.04} & \textbf{4.67} & \textbf{3.13} & \textbf{84.85} & 93.72\\
        \Xhline{1.2pt}
    \end{tabular}%
    }

    \vspace{4pt}
    \caption*{
    \small
    \textbf{Details:} I-QSPR 1, I-QSPR 2, I-QSPR 3: Intermediate models using data-driven features. 
    E-QSPR: Expert-curated model. 
    QSPR: Final model combining data-driven and expert-curated features. 
    In the absence of expert features, I-QSPR 3 serves as the final QSPR.\\
    AUC: area-under-the-curve features from spectra and its second derivative; 
    $p$: processing conditions; 
    $\sigma$: conductivity; 
    $M$: interaction products between AUC features; 
    $D$: SHAP-selected data-driven subset of AUC, $p$, and $M$; 
    $E$: expert-identified features; 
    $C$: SHAP-selected best subset from $D$ and $E$.}

\end{table}

\subsection{Domain-Knowledge Based Feature Expansion - Intermediate QSPR Model 2}

\begin{figure*}[t!]
	\centering
\includegraphics[width=1\linewidth]{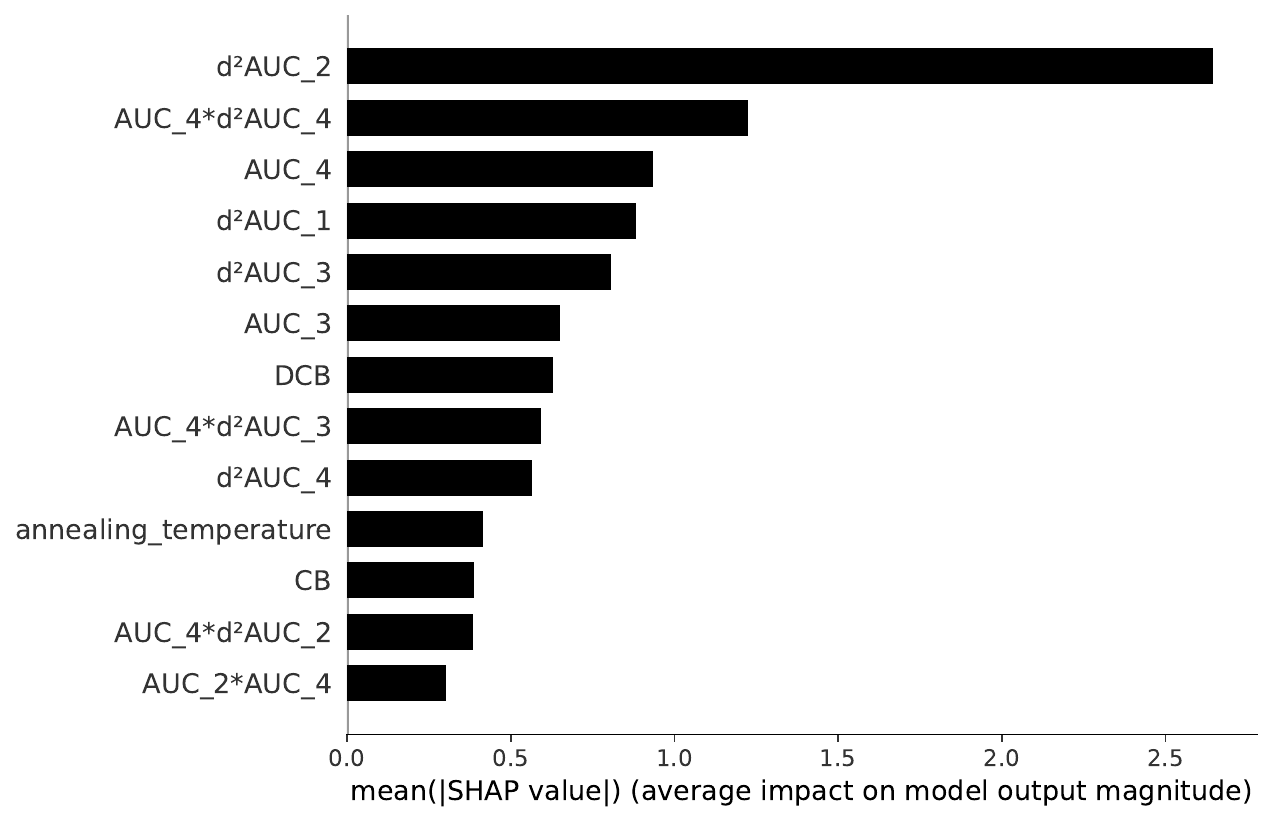}
    \caption{Feature importance (SHAP score) for each feature in I-QSPR model 2 (13 features which gave the best I-QSPR model 3 shown)}
    \label{Fig:shap_bar}
\end{figure*}
To further improve model performance, we expanded the feature set by applying simple mathematical transformations to the AUC features. Mathematical transformations, such as ratios, products, logarithms, and exponentials, could be applied to the AUC features. While a wide range of transformations could theoretically be explored, unrestricted application of all combinations would lead to a combinatorial explosion in the number of features, increasing the risk of overfitting.

In our case, the selection of mathematical transformations was guided by domain knowledge. Product and ratio transformations between the AUC features were identified as meaningful. It captured the underlying physical interactions between spectral regions that influence conductivity. These derived features could be used to improve the model’s predictive capability. We tested both the ratio and product mathematical transformations. We observed that for our problem, the product gave us slightly better performance compared to the ratio.

We computed the pairwise product of all combinations of AUC features. With five bin locations, this resulted in 8 primary AUC features (from the original and second-derivative spectra) and 28 interaction features (8 choose 2, $\binom{8}{2}$), in addition to the 4 processing condition features, yielding a total of 40 input features.

We trained another ML model using this expanded feature set. We call this model the intermediate QSPR model 2.  However, as shown in Table~\ref{tab:model1_results}, the model’s performance on the test set was similar to I-QSPR model 1. The likely reason is overfitting due to the high dimensionality of the feature space relative to the dataset size ~\cite{hua2005optimal}. The inclusion of many correlated features, especially those from the AUCs of both the original and second-derivative spectra, as well as their products, compromises generalization. Given this redundancy, feature selection becomes essential to remove irrelevant or correlated features. 

\subsection{SHAP-based Feature Selection}

\label{shap_forward}

For feature selection, tree-based ensemble models, such as Random Forest and Gradient Boosting, provide a built-in mechanism for estimating feature importance. These models build multiple decision trees using bootstrapped samples of the data and subsets of features. During training, features are selected at splits based on how well they reduce impurity (e.g., variance or Gini index). The total reduction in impurity contributed by each feature across all trees yields a global importance score.

However, tree-based feature importance has limitations. First, it is not model-agnostic. It relies on how a specific tree-based model splits the data during training. As a result, the importance scores reflect the internal structure and decision rules of that particular model, which can vary with different datasets or model configurations. Moreover, relying on tree-based methods for feature importance restricts us to tree-based models when building QSPRs. While such models performed well in our case, this may not always be the case. In certain scenarios, simpler models, such as linear regression, may offer better performance. Although linear models provide coefficients that can serve as indicators of feature importance, these can be misleading in the presence of multicollinearity or when feature scales vary. This limitation is partially addressed by LASSO regression, which applies L1 regularization to shrink irrelevant coefficients to zero, thereby enabling feature selection and enhancing interpretability. However, LASSO still assumes linear relationships and cannot capture interaction effects. Second, tree-based importance may also miss such interactions, where the relevance of one feature depends on another. Finally, these methods typically provide only global explanations, offering limited insight into individual predictions.

To address these limitations, we employ SHAP (SHapley Additive exPlanations)~\cite{lundberg2017unified}, a model-agnostic method based on cooperative game theory. SHAP computes the contribution of each feature to the prediction for each individual data point, offering both global and local interpretability. The SHAP framework represents the model output as an additive model. It is mathematically represented as:

\begin{equation}
f(x) = f_{baseline} + \sum_{i=1}^{M} \phi_i
\label{eq:shap_eq}
\end{equation}
where,

$f(x)$: Model prediction for given a input $x$,  

$f_{baseline}$: Average model prediction

$\phi_i$: SHAP value for feature $i$, indicating its contribution to $f(x)$

SHAP values are calculated as the average marginal contribution of a feature across all possible feature subsets:

\begin{equation}
    \phi_i = \sum_{S \subseteq N \setminus \{i\}} \frac{|S|! \, (M - |S| - 1)!}{M!} \left[ f_{S \cup \{i\}}(x) - f_S(x) \right]
\label{eq:shap_phi}
\end{equation}
where,

$M$: Total number of features

$N = {1,2, ... ,M}$: Set of all feature indices

$i \in N$: The index of the feature we are computing the SHAP value for

$S \subseteq N \setminus \{i\}$: A subset of all features excluding feature $i$

$f_S(x)$: Expected model output when only features in set $S$ are known

$f_{S \cup \{i\}}(x)$: Expected model output when feature $i$ is added to subset $S$

\smallskip
$\frac{|S|! \, (M - |S| - 1)!}{M!}$  is the Shapley weight and represents the probability of a particular subset $S \subseteq N \setminus \{i\}$ appearing before feature $i$ in a random ordering of all features. This weight ensures that all possible feature orderings are fairly considered when computing the contribution of feature $i$. $f_{S \cup \{i\}}(x) - f_S(x)$, measures the marginal contribution of feature  $i$ when added to subset $S$. It quantifies how much the prediction changes when feature $i$ is included, compared to using only the features in $S$. This captures the added value of feature $i$ given the context of subset $S$. SHAP provides the average marginal contribution of each feature across all possible subset of features. It also guarantees mathematical properties, specifically, a) efficiency: the sum of contributions of all features equals the difference between total prediction and average prediction, b) symmetry: features that equally contribute have equal SHAP values, c) zero contribution: if a feature does not affect the prediction, its SHAP value is zero, and d) linearity: if two models are combined, the SHAP value for a feature in the combined model is equal to the sum of it's SHAP value in each individual model. SHAP provides an importance ranking for each feature based on its average contribution to the model.

\begin{figure}[htbp]
	\centering
\includegraphics[width=1\linewidth]{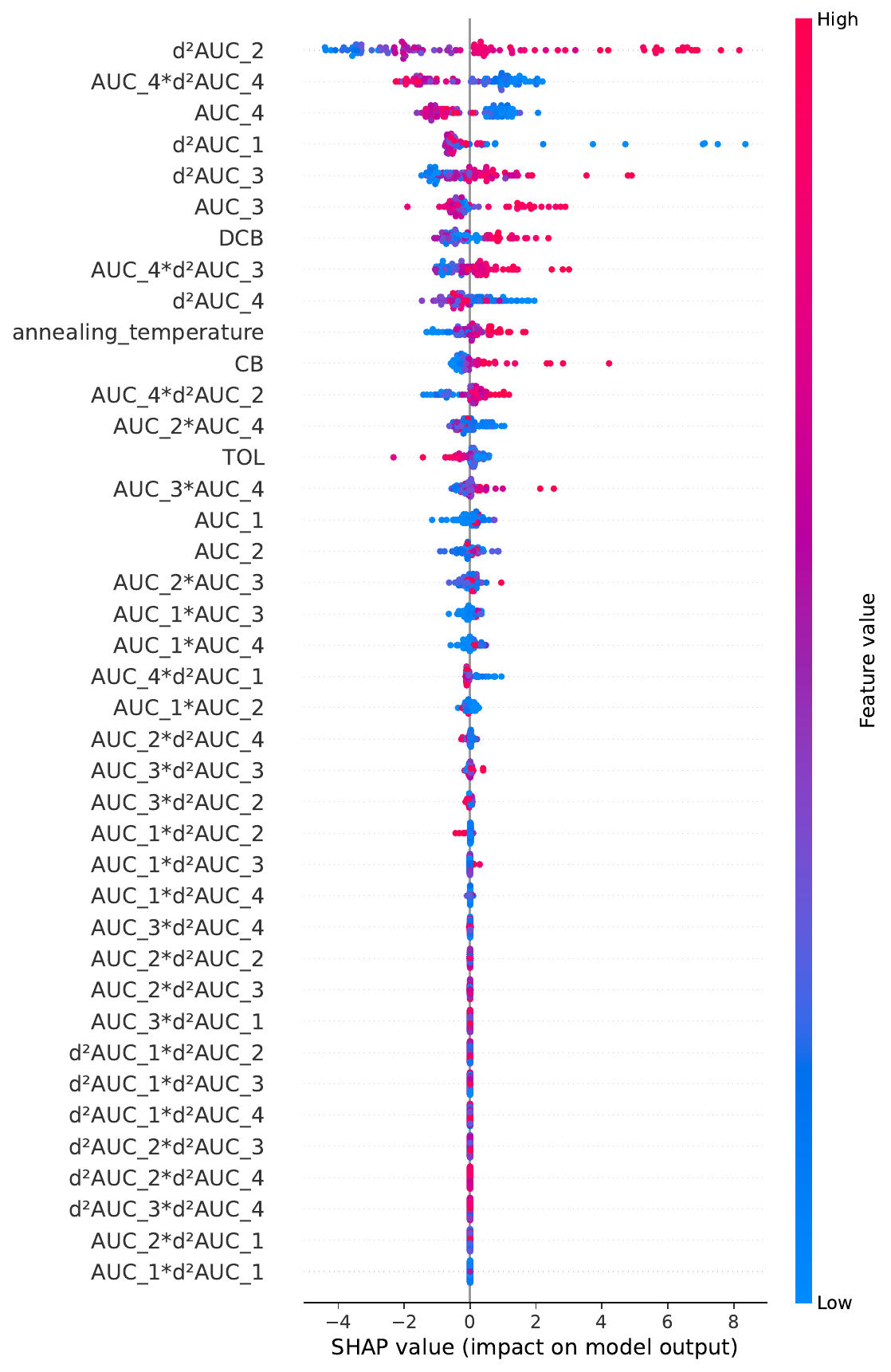}
    \caption{SHAP score for each sample in test dataset showing directional SHAP score for each feature in I-QSPR 2}
    \label{Fig:shap_beeswarm_all}
\end{figure}

We compute the mean absolute SHAP score for each input feature to evaluate its contribution to the model’s predictions. We chose the random forest algorithm-based model obtained from I-QSPR 2 as it gave the best performance compared to other algorithms (Table \ref{tab:all_model_results}). We only use the training dataset to rank the features. Table \ref{tab:feature_descriptions} lists the key for all the features, and Figure~\ref{Fig:shap_bar} shows the SHAP scores for the most important subset of features. To provide instance-level interpretability, Figure~\ref{Fig:shap_beeswarm_all} (Appendix \ref{appendix5}) presents SHAP scores across individual training samples, highlighting how each feature helps in conductivity prediction relative to the model’s mean prediction. SHAP is used here to rank feature importance and to support feature selection within the trained QSPR models, rather than to infer causality. Accordingly, the SHAP-based analysis is interpreted in conjunction with established physical understanding, and no causal claims are made.

\begin{figure*}[t!]
	\centering
\includegraphics[width=0.5\linewidth]{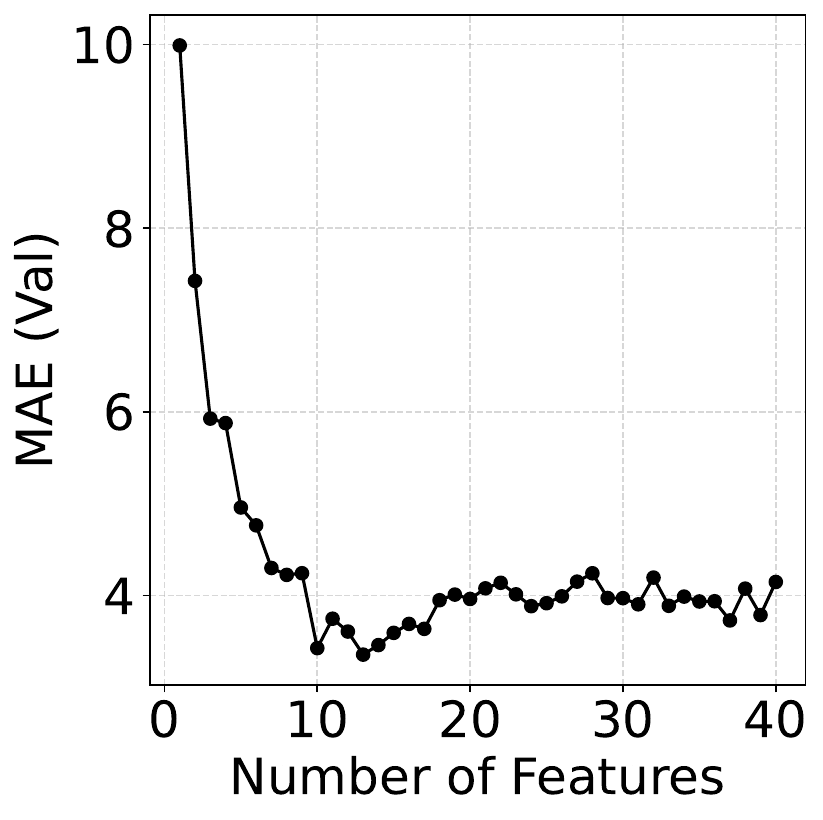}
    \caption{Mean Absolute Error (MAE) for models trained by starting with the most important feature and then subsequently adding important features identified by SHAP to the feature set and training a new model. The maximum validation MAE is obtained by a model with 13 features. This model is I-QSPR 3. Note that models are trained only on the train dataset, and this plot shows performance on the validation set. The final model with 13 features is further evaluated on the unseen test set.}
    \label{Fig:r2testfs_3}
\end{figure*}

To identify the most important features, we use a SHAP-guided greedy forward selection strategy. Features are added one by one according to their SHAP importance ranking. At each step, models are trained on the training data and evaluated on the validation set, and the feature subset that minimizes the mean absolute error (MAE) is selected. Ties are resolved using the root-mean-square error (RMSE). MAE is chosen as the primary selection metric because it is more robust in small-dataset settings, where individual data points have a large influence on evaluation metrics. In our study, the validation and test sets contain only 13 and 12 samples, respectively, meaning that a single data point represents approximately 8–9\% of the dataset. In the presence of outliers, both $R^2$ and RMSE can vary strongly and lead to unstable feature selection. In contrast, MAE penalizes errors linearly, providing a more stable and reliable basis for model comparison. This approach allows us to identify a compact set of informative features that improves generalization while removing redundant or highly correlated features that do not contribute additional predictive value. Figure \ref{Fig:r2testfs_3} shows the validation MAE for the 40 trained models. We observe that the model with 13 features achieves the minimum MAE in the validation set. These 13 features are:

\begin{enumerate}
  \item $d^{2}\mathrm{AUC}\_{2}$: AUC for the second derivative of optical spectra between  \texttt{(1.828, 1.982)}  eV.
  \item $\mathrm{AUC}\_{4}*d^{2}\mathrm{AUC}\_{4}$: Product of AUC for original spectra between \texttt{(2.095, 2.700)} eV and AUC for the second derivative of optical spectra between \texttt{(2.095, 2.700)} eV.
  \item $\mathrm{AUC}\_{4}$: AUC of the optical spectra between \texttt{(2.095, 2.700)} eV.
  \item $d^{2}\mathrm{AUC}\_{1}$: AUC for the second derivative of optical spectra between \texttt{(1.378, 1.828)}  eV
  \item $d^{2}\mathrm{AUC}\_{3}$: AUC for the second derivative of optical spectra between  \texttt{(1.982, 2.095)} eV.
  \item $\mathrm{AUC}\_{3}$: AUC of the optical spectra between \texttt{(1.982, 2095)} eV.
  \item DCB: Ortho-dichlorobenzene volume fraction (\%).
  \item $\mathrm{AUC}\_{4}*d^{2}\mathrm{AUC}\_{3}$: Product of AUC for original spectra between \texttt{(2.095, 2.700)} eV and AUC for the second derivative of optical spectra between \texttt{(1.982, 2.095)} eV.
  \item $d^{2}\mathrm{AUC}\_{4}$: AUC for the second derivative of optical spectra between \texttt{(2.095, 2.700)}  eV
  \item annealing\_temperature: Annealing temperature (°C).
  \item CB: Chlorobenzene volume fraction (\%).
  \item $\mathrm{AUC}\_{4}*d^{2}\mathrm{AUC}\_{2}$: Product of AUC for original spectra between (2.095, 2.700) eV and AUC for the second derivative of optical spectra between \texttt{(1.828, 1.982)} eV
  \item $\mathrm{AUC}\_{2}*\mathrm{AUC}\_{4}$: Product of AUC for original spectra between \texttt{(1.828, 1.982)} eV and \texttt{(2.095, 2.700)} eV
\end{enumerate}
Readers are also referred to Table \ref{tab:feature_descriptions} for descriptions of the features.

\subsubsection{Intermediate QSPR Model 3}

Using the identified important features, we train a regression model, referred to as intermediate QSPR Model 3. This model improves the test $R^2$ by approximately 3\% over the I-QSPR model 1, as shown in Table~\ref{tab:model1_results}. It also outperforms the I-QSPR Model 1 across other evaluation metrics, including RMSE, MAE, and Pearson correlation. These results demonstrate that combining domain-knowledge-based feature expansion with data-driven feature engineering enhances overall model performance.

I-QSPR Model 3 can serve as a surrogate for direct conductivity measurements. As shown in Figure~\ref{Fig:time_analysis}, the conductivity measurement accounts for roughly 33\% of the total experimental time. By replacing it with model predictions, we can significantly reduce the experimental burden, thereby enabling higher-throughput experimentation. Moreover, in our current experimental workflow, the post-anneal spectrum is found to be the most informative. Therefore, for studies focused solely on polymer processing, theoretically, an experimental time reduction of up to 50\% can be achieved by omitting post-doping steps. However, this simplification is only applicable when post-doping spectra do not provide additional relevant information. Next, we train a new model based on expert-identified features to compare against the results obtained from data-driven features.

\subsection{Conductivity Prediction Using Expert Features - E-QSPR}

\begin{figure*}[t!]
	\centering
\includegraphics[width=1\linewidth]{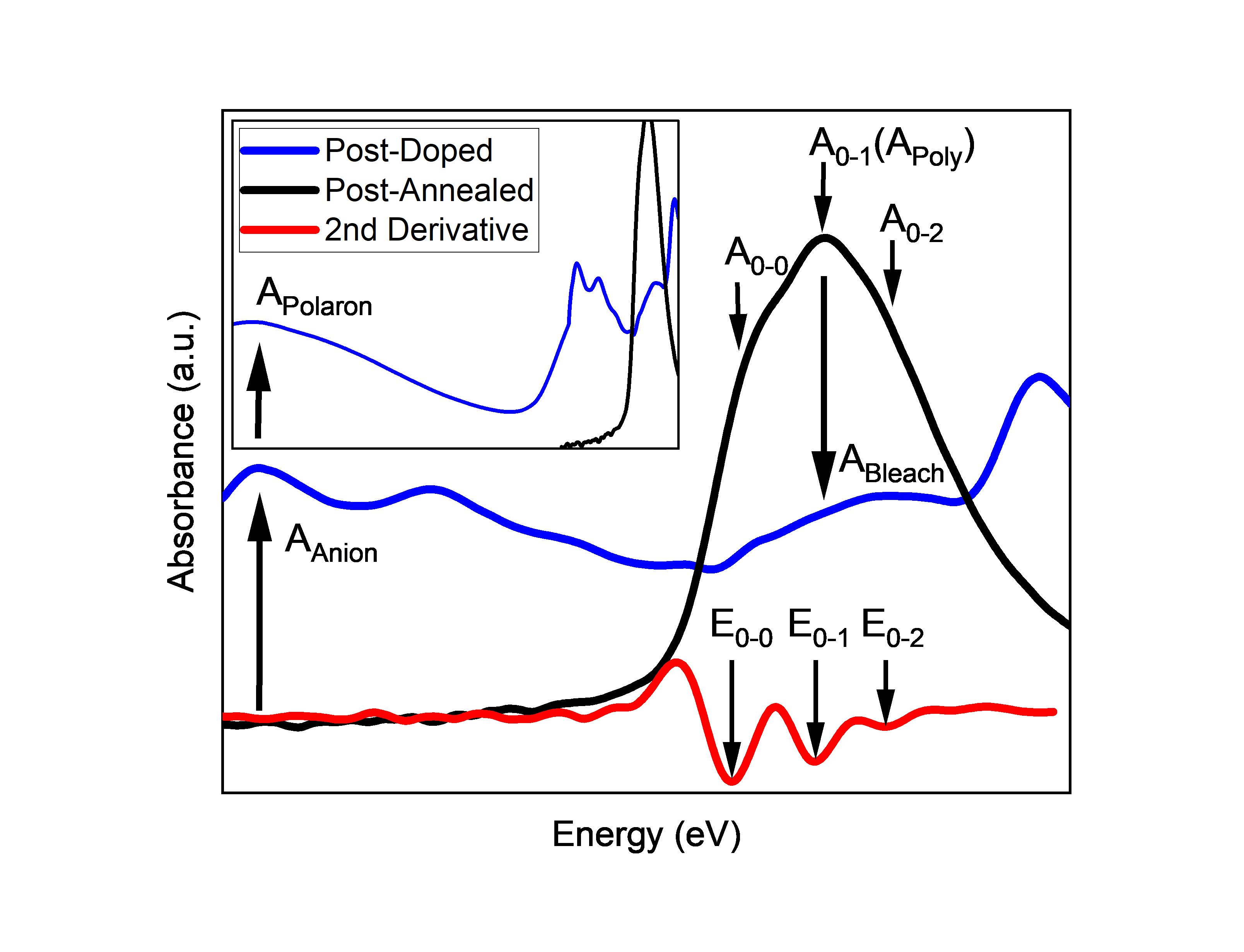}
    \caption{Expert-identified features were derived through an extensive literature review and validated using experimentally collected data. These features exhibit strong correlation with conductivity and represent the outcome of over a year of analysis. A detailed account of the feature identification process is provided in a separate publication by our team.}
    \label{Fig:expert_features}
\end{figure*}

In our related work \cite{Mauthe2026AI}, seven spectral features were identified by domain experts through an extensive literature review and validation using experimentally collected data. This effort, which involved a literature survey, prior knowledge of the conjugated polymer, and generation of spectral data from 128 individual samples, resulted in a set of features highly correlated with electrical conductivity. Over the course of one year, our companion work identified features originating from the annealed and doped spectroscopy, along with other characterization techniques not included in this study. The identified features are illustrated in Figure~\ref{Fig:expert_features} and include:

\begin{itemize}
\item $E_{0-0}$: Energy corresponding to the zeroth valley in the second derivative of the post-annealed spectrum.
\item $E_{0-1}$: Energy corresponding to the first valley in the second derivative of the post-annealed spectrum.
\item $E_{0-2}$: Energy corresponding to the second valley in the second derivative of the post-annealed spectrum.
\item $A_{0-0}/A_{0-1}$: Ratio of absorbance values at $E_{0-0}$ and $E_{0-1}$.
\item \textit{\% Bleaching}: Ratio of $A_{\text{Bleach}}$ (post-dope spectrum) to $A_{0-1}$ ($A_{\text{poly}}$, post-anneal spectrum).
\item \textit{Anion Signal}: Ratio of $A_{\text{Anion}}$ to $A_{\text{Bleach}}$.
\item \textit{Polaron Signal}: Ratio of $A_{\text{Polaron}}$ to $A_{\text{Bleach}}$.
\end{itemize}

These features are described in detail in our companion publication \cite{Mauthe2026AI}. We trained a machine learning model using these expert-curated features (referred to as E-QSPR). The model's performance was found to be slightly better than that of I-QSPR Model 3, as shown in Table \ref{tab:model1_results}. 

This result highlights the effectiveness of our data-driven feature extraction strategy, which systematically identifies informative spectral regions using AUC combined with GA. These features, when further refined through expert-guided transformations and feature engineering, achieve predictive performance comparable to that of expert-identified features. Importantly, our approach is more efficient because optimal bin selection and model training can be completed within a few hours. This demonstrates the potential of our hybrid strategy, which combines domain knowledge with automated feature discovery, as a scalable alternative to traditional expert-driven analysis, which is time-consuming.

\subsection{Combining Data-Driven Features and Expert-Identified Features - Final QSPR Model}

\begin{figure*}[t!]
	\centering
\includegraphics[width=1\linewidth]{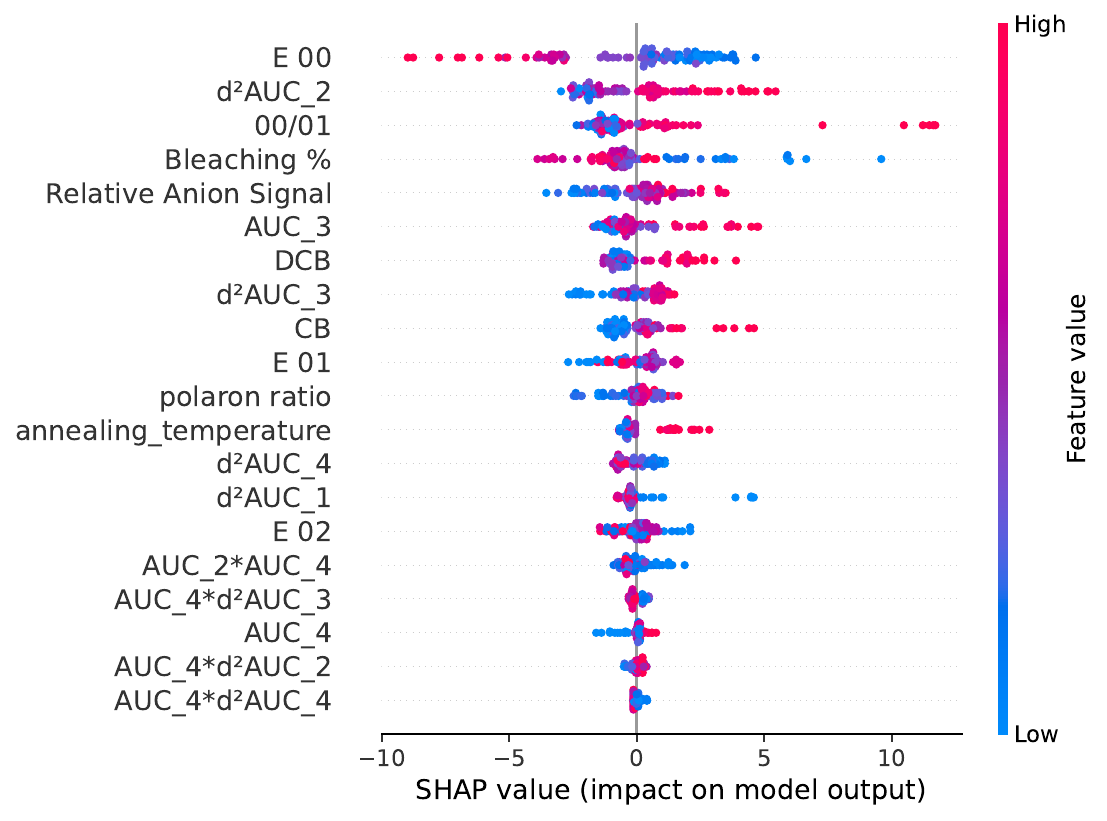}
    \caption{SHAP score for each sample showing directional SHAP score for data-driven features and expert-identified features}
    \label{Fig:shap_beeswarm_QSPR5}
\end{figure*}

\begin{figure*}[t!]
	\centering
\includegraphics[width=0.5\linewidth]{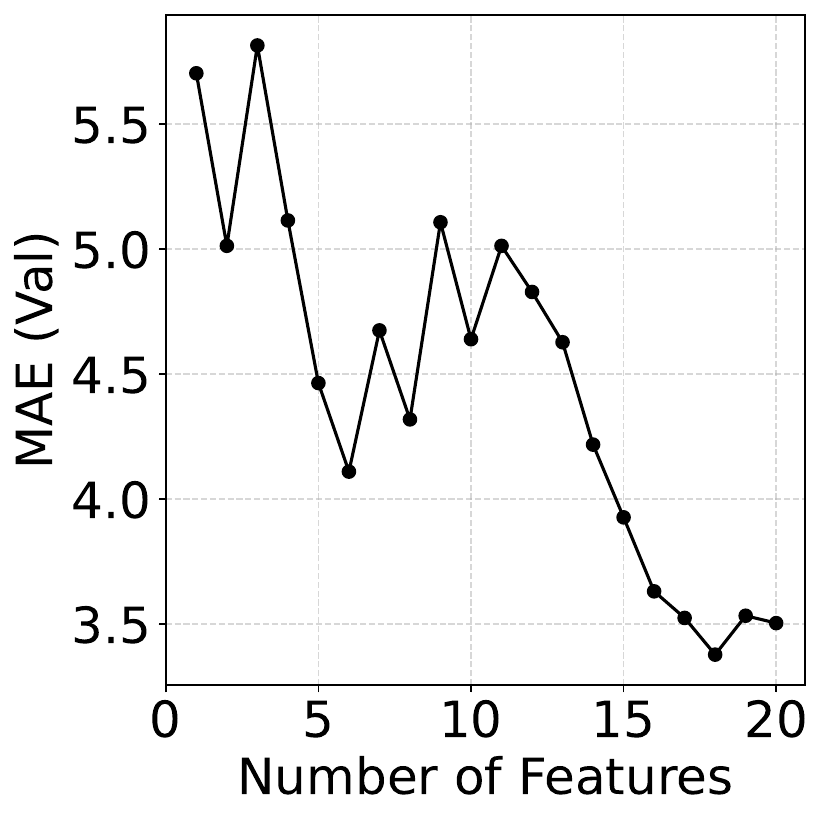}
    \caption{Mean absolute Error (MAE) of validation set for 20 models trained by starting with the most important feature and then subsequently adding important features identified by SHAP to the feature set and training a new model. We use 13 data-driven features and 7 expert-identified features. The minimum MAE is obtained by a model with 18 features. This model is the final QSPR. We further evaluate the model on the unseen test set.}
    \label{Fig:r2testfs}
\end{figure*}

\begin{figure*}[t!]
	\centering
\includegraphics[width=1\linewidth]{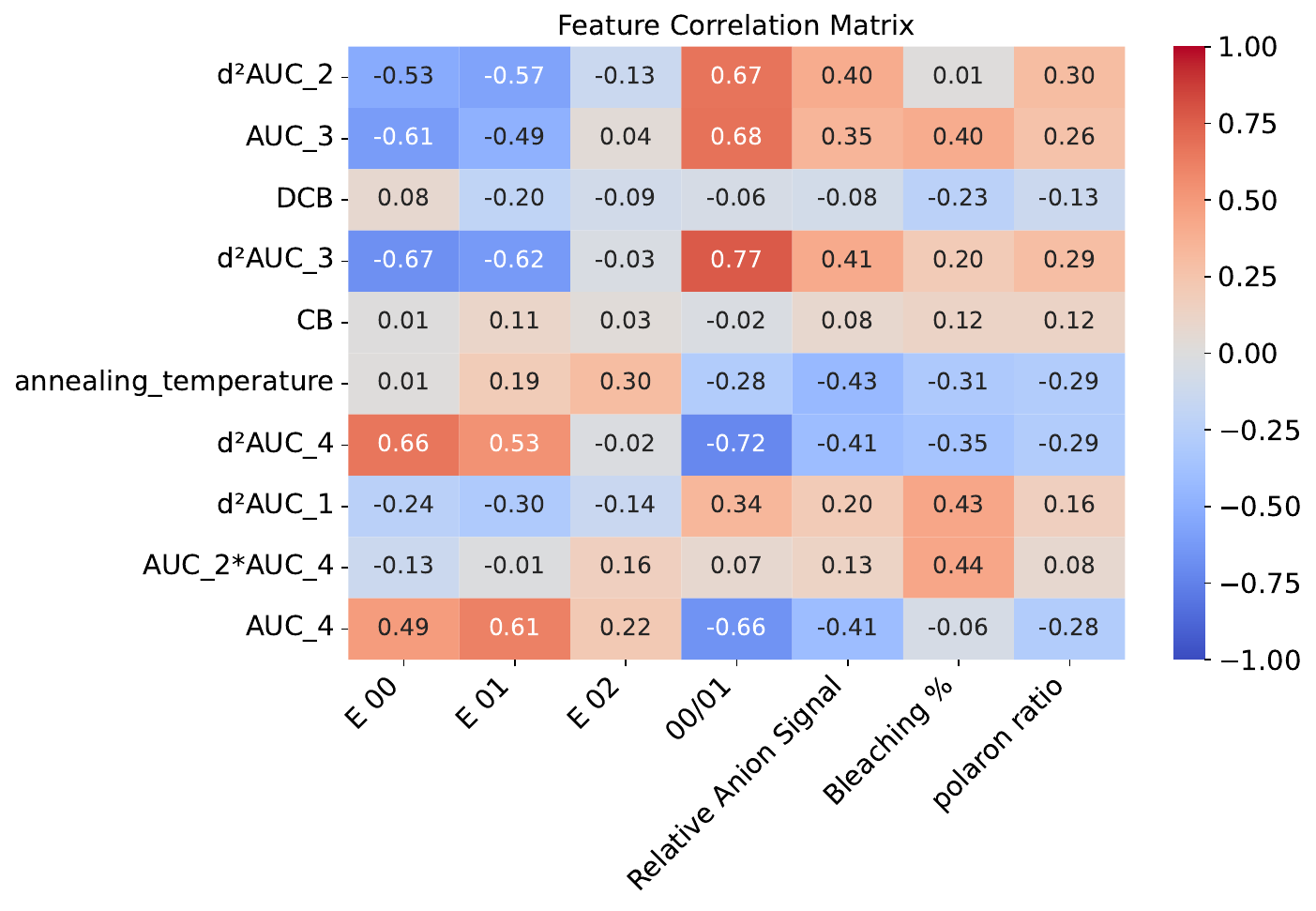}
    \caption{Spearman correlation between data-driven features (y-axis) and expert-curated features (x-axis) for final QSPR}
    \label{Fig:correlation_specific}
\end{figure*}

We combine the data-driven features (13 in total) with expert-identified features (7 in total) to examine whether integrating expert knowledge with machine learning leads to improved model performance. A SHAP analysis is conducted to evaluate the importance of each feature, as shown in Figure \ref{Fig:shap_beeswarm_QSPR5}. Guided by the SHAP-based ranking, we apply a greedy forward-selection strategy, described in Section \ref{shap_forward}, to identify the most informative subset of features and the corresponding best-performing model. Figure \ref{Fig:r2testfs} shows the validation MAE for all 20 models.  The minimum MAE on the validation set is achieved using 18 features.  We also observe that the feature $AUC\_4*d^2AUC\_3$ has a perfect correlation with $ d^2AUC\_3$. So, we drop the feature $AUC\_4*d^2AUC\_3$.  We then further evaluate the model with the 17 features on the test set. We achieve an $R^2$ of 85\% on the test set. This represents an improvement of approximately $\sim$9\% compared to the model built using only data-driven features and $\sim$4\% compared to the model only using expert-identified features, highlighting the potential of combining human expertise with machine learning. Among the 17 selected features, 7 were expert-curated, and 10 were data-driven. Of the data-driven features, three corresponded to processing conditions, while the remaining seven were derived from AUC-based spectral features. A feature correlation matrix illustrating the relationship between data-driven and expert features is provided in Figure~\ref{Fig:correlation_specific}. 

Below, we provide a brief analysis of the 7 data-driven spectral features from the combined final QSPR  and their connection to the expert-identified features:

$d^{2}\mathrm{AUC}\_{2}$: AUC of the second derivative of optical spectra between \texttt{(1.828,\,1.982)} eV. This feature captures the initial maximum in the second derivative spectrum, which comes from the polymer 0-0 peak onset. A high value corresponds to a red-shifted E0-0, indicative of higher aggregation, which leads to higher conductivity. This is reinforced by the strong correlations of this feature with the E0-0 and E-01 energies, as well as 0-0/0-1 peak ratio, as shown in Figure \ref{Fig:correlation_specific}. 

$\mathrm{AUC}\_{3}$: AUC of the optical spectra between \texttt{(1.982, 2.095)} eV. The area under the curve of this region directly reflects the prominence of the 0-0 vibronic transition relative to the other spectral regions, as well as the width/broadness of the peak onset. In pBTTT films with higher aggregation, this 0-0 peak should be more prominent; this increased aggregation tends to lead to higher mobility and thus conductivity after doping. This is confirmed by the strong correlations of this feature with the E0-0 and 0-0/0-1 ratio in Figure \ref{Fig:correlation_specific}. Interestingly, this feature is also correlated with the bleaching. This may indicate that lower energy 0-0 peaks result in a density of state more suitable for doping with F4TCNQ. This is further investigated in our companion work.

$d^{2}\mathrm{AUC}\_{3}$: AUC for the second derivative of optical spectra between  \texttt{(1.982, 2.095)} eV. This feature captures the peak position of the 0-0 vibronic transition, a deep local minimum in the second derivative (leading to higher values in the SHAP analysis, Figure \ref{Fig:shap_beeswarm_QSPR5}), indicating the strength and sharpness of the 0-0 transition. This is closely tied to the order and aggregation of the polymer, as evident in the SHAP analysis, which shows high values leading to improvements in the estimated conductivity. This is reinforced by the very strong correlations of this feature with the E0-0 and E-01 energies as well as 0-0/0-1 peak ratio as noted in Figure \ref{Fig:correlation_specific}.

$d^{2}\mathrm{AUC}\_{4}$: AUC for the second derivative of optical spectra between \texttt{(2.095, 2.700)}  eV. This spectral region captures the higher energy vibronic transitions (E0-1 \& E0-2). The local minima in the second derivative are conventionally used to identify these peak locations. The prominence of these minima indicates the intensity of these transitions relative to the 0-0 transition, as well as reflects the positioning of E0-1. A higher area under the curve would indicate strong 0-1 transitions, a sign of disorder and lowered aggregation in pBTTT, which would lead to decreases in conductivity. This is reinforced with the positive correlation with E0-0 and E0-1 as well as the negative correlation with the 0-0/0-1 ratio shown in Figure \ref{Fig:correlation_specific}.

$d^{2}\mathrm{AUC}\_{1}$: AUC for the second derivative of optical spectra between \texttt{(1.378, 1.828)}  eV. This spectral region captures the low-energy tail states. These low-energy tail or trap states are typically found in the amorphous regions of the film and often serve as the initial doping sites. The SHAP analysis in Figure \ref{Fig:shap_beeswarm_QSPR5} indicates that a few samples with very low values in this spectral region tend to have higher conductivity. This makes sense as the same amorphous regions that give rise to these trap states tend to have very low mobility, leading to overall lowered conductivity. This is also reinforced by the correlation with bleaching shown in Figure \ref{Fig:correlation_specific}. Notably, this feature is not correlated with any of the pre-doping spectroscopic features identified in our companion study \cite{Mauthe2026AI}.

$\mathrm{AUC}\_{2}$*$\mathrm{AUC}\_{4}$: The product of the AUC of the optical spectra for the \texttt{(1.828,\,1.982)} eV and \texttt{(2.095, 2.700)}  eV regions. The former region exists below the 0-0 transition and represents low-energy tail states. As previously noted these states often serve as initial doping sites in conjugated polymers though can often lead to lower mobility carriers. This is also reinforced by the correlation with bleaching shown in Figure \ref{Fig:correlation_specific} as well as samples with low feature value having a positive SHAP value in Figure \ref{Fig:shap_beeswarm_QSPR5}. The latter spectral region captures the higher energy vibronic transitions (E0-1 \& E0-2). The prominence of these transitions, particularly when considered relative to the prominence of the 0-0 transition, are a sign of heightened disorder or lowered aggregation in pBTTT, which would lead to decreases in conductivity as reinforced by the SHAP analysis. Based on the correlation analysis in Figure \ref{Fig:correlation_specific}, the component of this feature appears to be the tail states as seen with higher correlation with bleaching compared to the 0-0/0-1 peak ratio.

$\mathrm{AUC}\_{4}$: AUC of the optical spectra between \texttt{(2.095, 2.700)}  eV. This spectral region captures the higher energy 0-1 \& 0-2 transitions. As noted in the previous features, this region tends to indicate enhanced disorder of the polymer when the value is high relative to the region containing the 0-0 transition. Though there is little impact in the model from low SHAP values seen in Figure \ref{Fig:shap_beeswarm_QSPR5}, the correlation analysis in Figure \ref{Fig:correlation_specific} indicates that this feature is indeed negatively correlated with physical features associated with aggregation, such as the 0-0/0-1 peak ratio.

\begin{figure*}[ht]
    \centering
    \begin{subfigure}[b]{0.32\textwidth}
        \includegraphics[width=\textwidth]{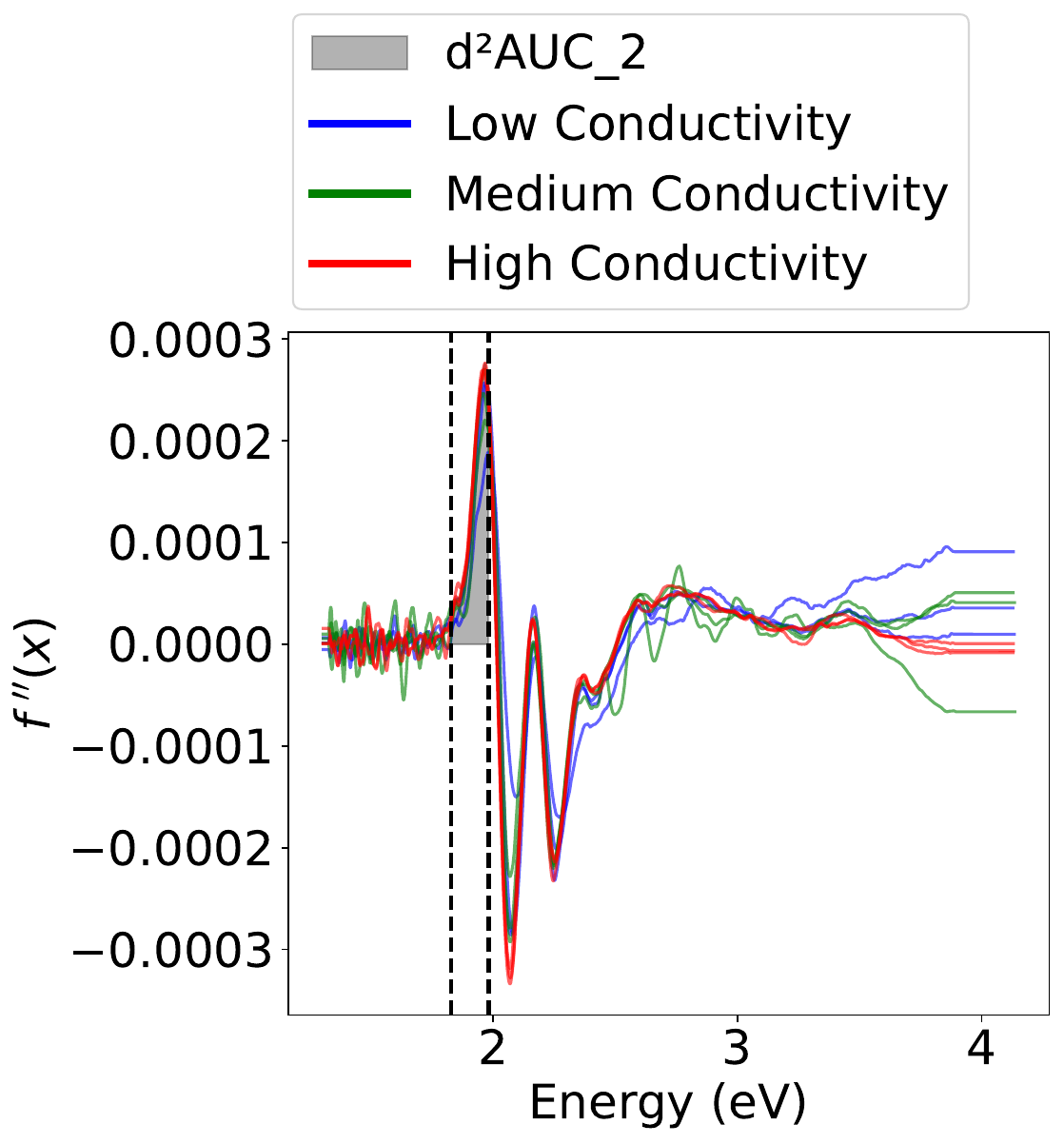}
        \caption{}
        \label{fig:sub1}
    \end{subfigure}
    \hfill
    \begin{subfigure}[b]{0.32\textwidth}
        \includegraphics[width=\textwidth]{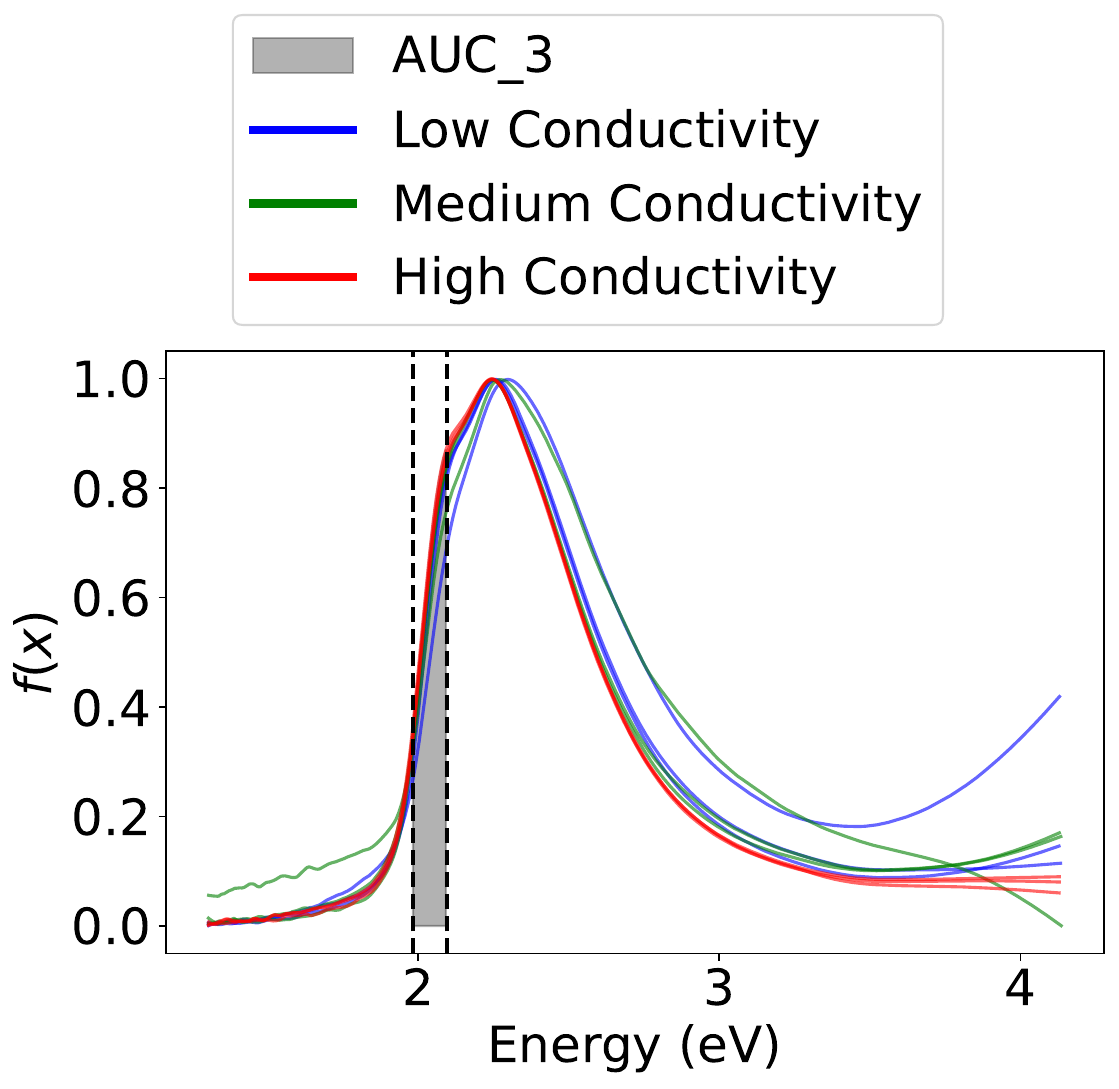}
        \caption{}
        \label{fig:sub2}
    \end{subfigure}
    \hfill
    \begin{subfigure}[b]{0.32\textwidth}
        \includegraphics[width=\textwidth]{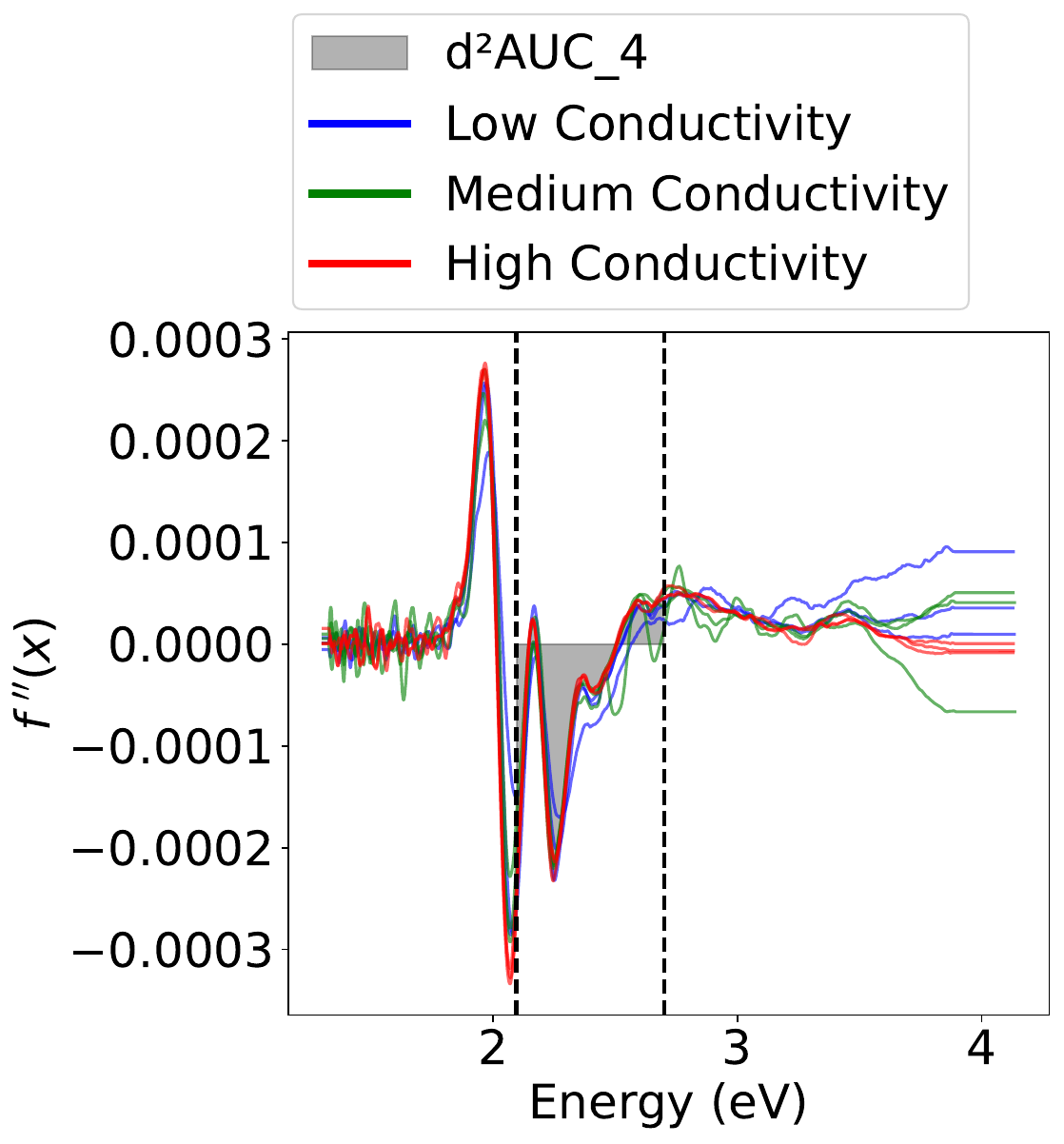}
        \caption{}
        \label{fig:sub3}
    \end{subfigure}

    \vspace{1em} 
    \begin{subfigure}[b]{0.32\textwidth}
        \includegraphics[width=\textwidth]{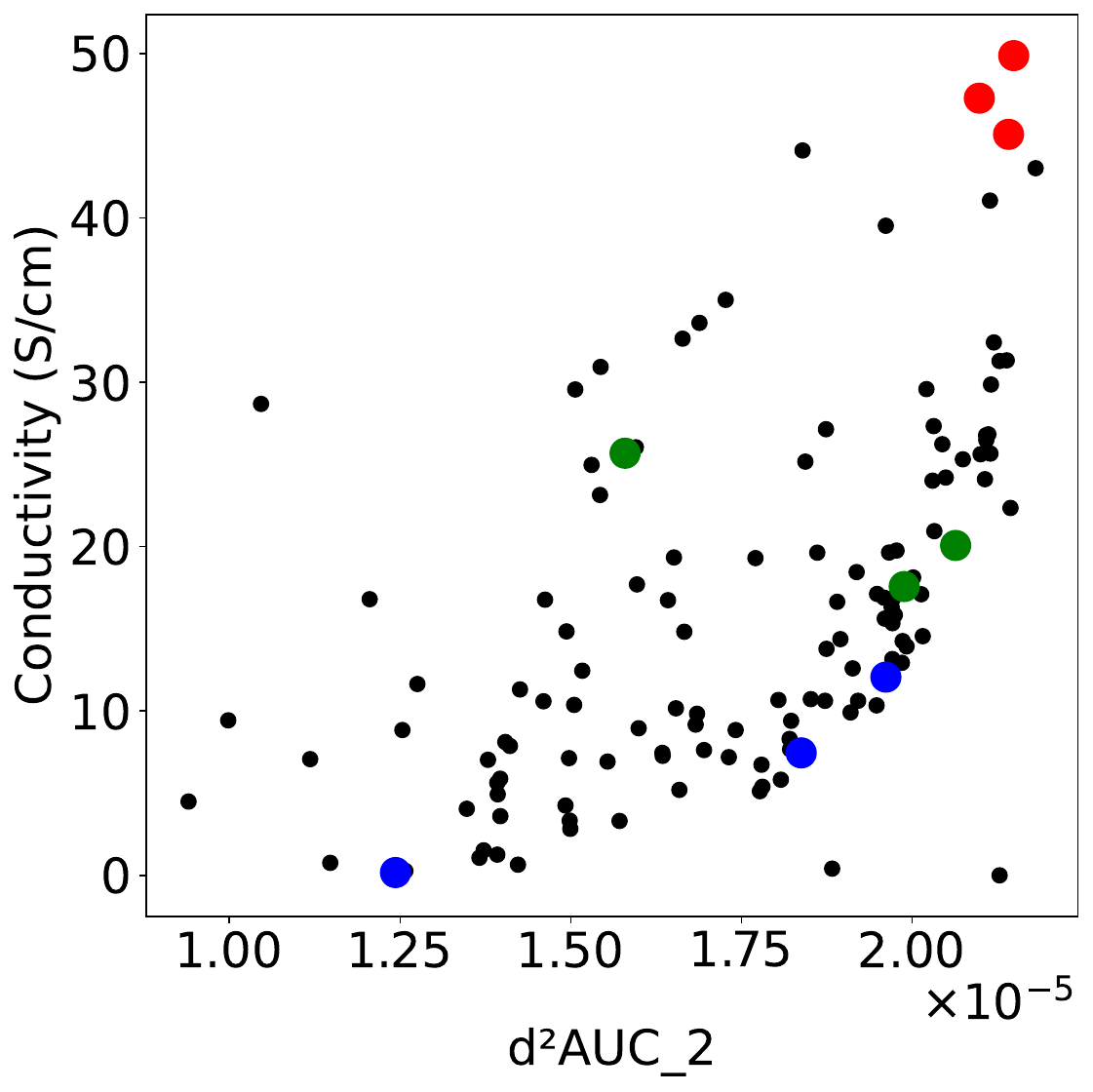}
        \caption{}
        \label{fig:sub4}
    \end{subfigure}
    \hfill
    \begin{subfigure}[b]{0.32\textwidth}
        \includegraphics[width=\textwidth]{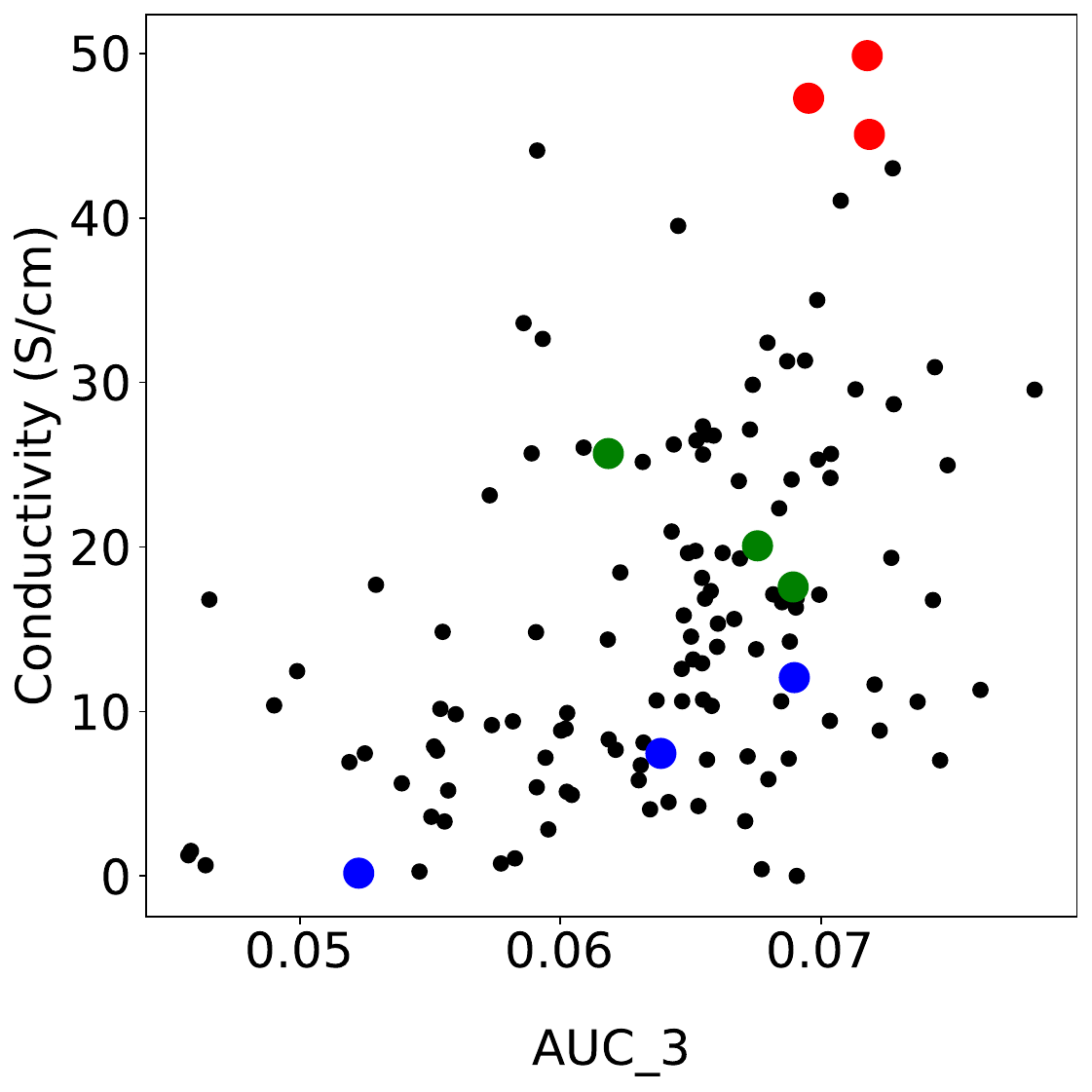}
        \caption{}
        \label{fig:sub5}
    \end{subfigure}
    \hfill
    \begin{subfigure}[b]{0.32\textwidth}
        \includegraphics[width=\textwidth]{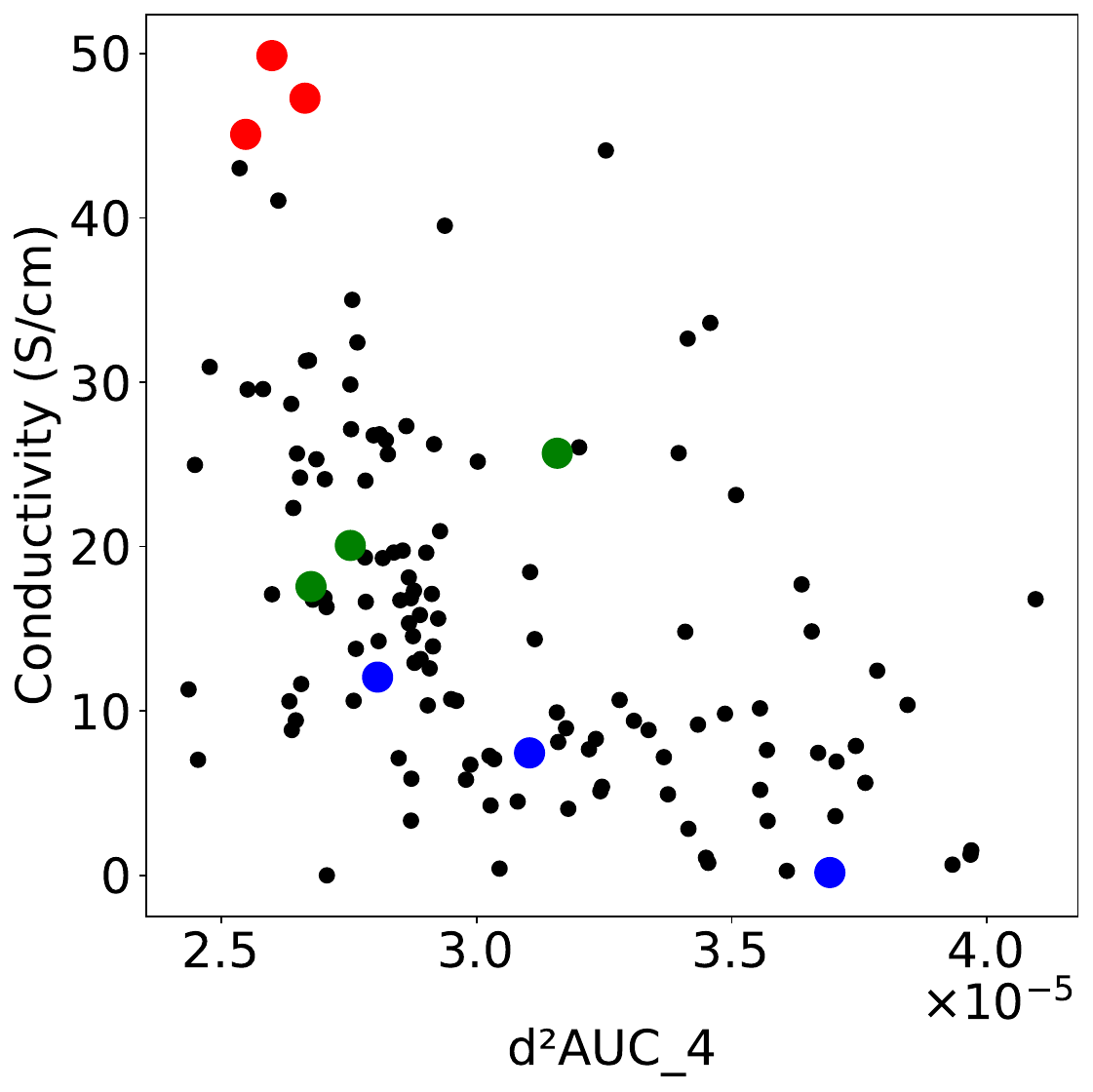}
        \caption{}
        \label{fig:sub6}
    \end{subfigure}

    \caption{Three representative samples each from the low (<16 S/cm), medium (16–32 S/cm), and high (32–50 S/cm) conductivity groups (total nine samples).  
(a) Second‐derivative spectra with the derivative feature region \texttt{1.8284–1.9825 eV} corresponding to feature $d^{2}\mathrm{AUC}\_{2}$ highlighted.  
(b) Original absorbance spectra with the feature region \texttt{1.9825–2.0952} eV corresponding to feature $\mathrm{AUC}\_{3}$ highlighted.  
(c) Second‐derivative spectra with the derivative feature region \texttt{2.0952–2.7003} eV corresponding to feature $d^{2}\mathrm{AUC}\_{4}$ highlighted.  
(d) Conductivity versus $d^{2}\mathrm{AUC}\_{2}$ feature (Pearson correlation = 52.29\%).  
(e) Conductivity versus $\mathrm{AUC}\_{3}$ feature (Pearson Correlation = 43.36\%).  
(f) Conductivity versus $d^{2}\mathrm{AUC}\_{4}$ feature (Pearson Correlation = -48.37\%).}
    \label{fig:six_panel_results}
\end{figure*}

\begin{figure*}[t!]
	\centering
\includegraphics[width=0.4\linewidth]{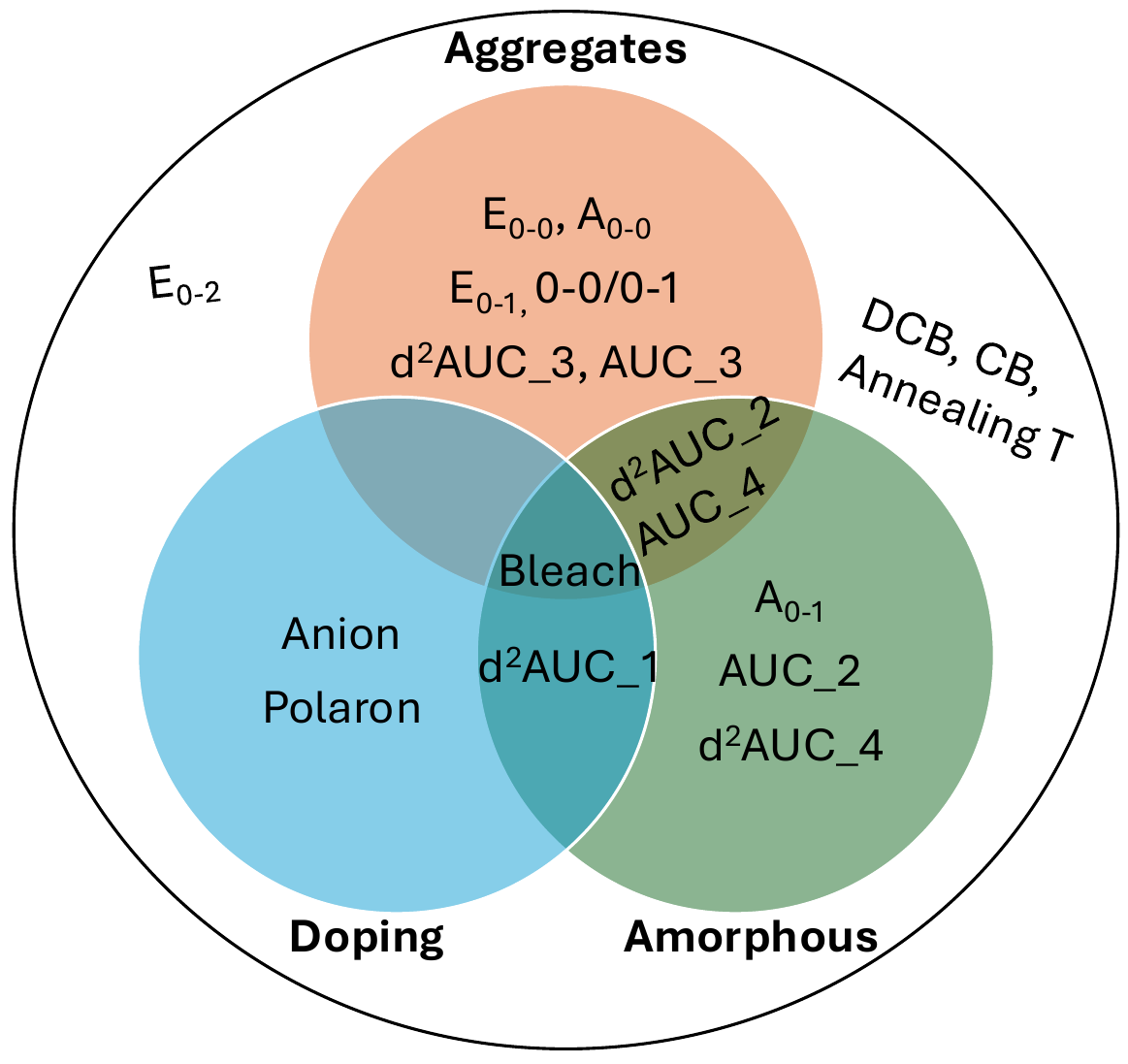}
    \caption{Venn diagram illustrating the overlap between data-driven features identified via spectral analysis and known materials descriptors related to aggregation, tail states, and doping phenomena. The convergence between machine-learned features (e.g., AUC and second-derivative features) and physically meaningful descriptors (e.g., aggregates, tail states, and doping signatures) underscores the interpretability and physical relevance of the proposed data-driven approach.}
    \label{Fig:venn_diagram}
\end{figure*}

The overall workflow described in the paper is shown in Figure~\ref{Fig:graph_abstract}. The process begins with spectral featurization using AUC combined with GA. Example graphs of the spectral featurization and of high, medium, and low conductivity samples are provided in Figure~\ref{fig:six_panel_results}. Following the data-driven featurization, domain knowledge-based features are incorporated, followed by feature engineering. Introducing additional features through simple, domain-informed mathematical operations, along with feature selection, leads to improved model performance. Further enhancement is achieved by integrating expert-curated features and refining the model, ultimately yielding the best-performing model. There is noticeable overlap in the data-driven features identified using this approach and the known materials descriptors for aggregation, tail states, and doping phenomena as highlighted in Figure~\ref{Fig:venn_diagram}. The improvement in model performance upon combining data-driven and expert-curated features demonstrates the value of synergizing human expertise with machine learning.

\section{Discussion}

In this work, we present a data-driven framework for feature extraction from optical spectra and prediction of electrical conductivity in doped conjugated polymers. Our approach combines area-under-the-curve (AUC) features with a genetic algorithm (GA) to automatically identify informative spectral regions. The resulting QSPR model, which is augmented with domain-knowledge transformations and targeted feature engineering, achieves predictive performance comparable to an expert-curated model while reducing time and manual effort.

Notably, the expert-curated features used here reflect an extensive literature review, domain insight, and manual validation, requiring roughly a year of dedicated effort. By contrast, the automated feature-extraction and model-training pipeline can be executed within hours, enabling rapid, scalable characterization. Because the model provides early conductivity predictions directly from spectra, it functions as a surrogate for direct conductivity measurements, reducing experimental time by approximately one-third and increasing throughput. Additional gains may be possible by broadening the library of transformations and automating their composition via systematic search and optimization.

Individually, the data-driven and expert-guided models exhibit similar performance; combining them yields a hybrid model with an $R^2$ of ~85\%, outperforming either model alone. This result highlights the value of human–AI synergy, where domain expertise and machine learning work together to deliver more accurate and interpretable predictors.

The framework also integrates naturally with multi-fidelity (Bayesian) optimization, where the QSPR acts as a low-fidelity surrogate and costly conductivity measurements are reserved for high-value candidates. Such workflows enable efficient exploration of large design spaces and support high-throughput experimentation. Overall, the hybrid strategy of combining expert knowledge with automated, data-driven analysis provides a scalable approach to accelerate materials discovery. It is well-suited to deployment in self-driving laboratories and to navigating complex design spaces in organic electronics and beyond.

This study has several limitations. First, the dataset is relatively small. This affects model complexity and limits the use of extensive cross-validation or uncertainty quantification without making performance estimates unstable. Second, the framework is shown on one material system, pBTTT: F4TCNQ. While the methodology is general, model performance and chosen features may depend on specific characteristics of this system. Third, the analysis uses only one spectral method. The approach's effectiveness with other spectroscopic techniques has not been tested and can be explored in the future. Fourth, uncertainty estimates are not reported since the analysis is based on a single train/validation/test split, not repeated resampling. Finally, the reported decrease in experimental time is a theoretical estimate based on the current workflow and has not been confirmed through closed-loop autonomous experiments. The integration of the proposed workflow in a full self-driving lab setting is an important next step to be explored in the future.

\section*{Funding Declaration}
We acknowledge support from ONR, United States, under award N00014-23-1-2001. J.M. and A.A. also acknowledge NC State’s Data Science Academy for support toward the design and development of the materials acceleration platform used in this project. B.G. acknowledges partial support from NSF 2323716.

\section*{Data Availability}

The data supporting the findings of this study are available at \url{https://github.com/ankush-kumar-mishra/InSpecLearn4SDL/tree/main/Data}. Experimental metadata, including processing conditions (solvent volume fractions and annealing temperatures) and measured electrical conductivity, are provided in a master CSV file. Additionally, the corresponding optical absorbance spectra, captured at three distinct states: as-cast, post-annealed, and post-doped, are provided as 128x3 individual CSV files containing wavelength and intensity data.

\section*{Code Availability}

All source code required to reproduce the results, including the Genetic Algorithm for spectral featurization, the QSPR model training pipeline (Random Forest and Gradient Boosting), and SHAP-guided feature selection scripts, is available at \url{https://github.com/ankush-kumar-mishra/InSpecLearn4SDL/tree/main/Code}. A detailed README file explaining the functional of each script is provided in the repository.

\section*{Author Contributions}

\textbf{AKM:}  Methodology, Software, Data Curation, Formal Analysis, Visualization, Writing - Original Draft \textbf{JPM:} Investigation, Validation, Visualization, Writing - Original Draft. \textbf{NL:} Investigation. \textbf{AA:} Conceptualization, Resources, Supervision, Writing - Review and Editing. \textbf{BG:} Conceptualization, Resources, Supervision, Project Administration, Formal Analysis, Writing - Review and Editing.

\section*{Competing interests}

The authors declare no competing interests.

\bibliographystyle{unsrt}
\bibliography{bibliography}
\section{Appendix}
\label{appendix}

\subsection{Appendix 1: Processing Parameter Selection}
\label{appendix1}
\FloatBarrier

\begin{table}[H]
\centering
\caption{Table of compatible solvents for pBTTT from HSP calculations with selected solvents bolded }
\label{tab:HSP}
\begin{tabular}{cccccc}
\hline
Solvent &  $ \delta D({Mpa}^{1/2}) $ & $ \delta P({Mpa}^{1/2}) $ & $ \delta H({Mpa}^{1/2}) $ & Soluble & RED\\ 
\hline
Acetone	& 15.5 &10.4 &	7 &	0 &	2.986\\ 
Acetonitrile &	15.3 &	18	& 6.1 &	0 &	4.748 \\
1-Butanol	& 16 &	5.7 &	15.8 &	0 &	4.076 \\
\textbf{Chlorobenzene} &	19 &	4.3 &	2	&1 &	0.471 \\
Chloroform &	17.8 &	3.1 &	5.7 &	1	& 0.952\\
\textbf{o-Dichlorobenzene} &	19.2	& 6.3	& 3.3	& 1	& 0.993 \\
1,1,2,2-Tetrachloroethane &	18.8 &	5.1 &	5.3 &	1	&0.934 \\
Tetrahydrofuran (Thf)	&16.8 &	5.7 &	8	& 0 &	1.957\\
1,2,4-Trichlorobenzene &	20.2 &	4.2 &	3.2 & 1 &	0.987\\
o-Xylene &	17.8 &	1 &	3.1 &	1 &	0.753\\
Ethyl Acetate &	15.8 &	5.3 &	7.2 &	0	& 2.128\\
Mesitylene &	18 &	0.6 &	0.6 &	1 &	0.999\\
\textbf{Toluene}	&18	&1.4 &	2	& 1 &	0.626\\
Cyclohexane	&16.8 &	0 &	0.2 &	0 &	1.533\\
n-Butyl Acetate (nBA)	& 15.8 &	3.7 &	6.3 &	0 &	1.923 \\
\hline
\end{tabular}
\end{table}

\begin{table}[H]
\centering
\caption{Hansen solubility parameters for pBTTT and F4TCNQ}
\label{tab:hansen}
\begin{tabular}{ccccc}
\hline
Material & $\delta D({Mpa}^{1/2})$ & $\delta P({Mpa}^{1/2})$ & $\delta H({Mpa}^{1/2}) $ & $R_0$\\ 
\hline
pBTTT-C14 & 18.6 & 3.2  & 2.6 & 3.5 \\ 
F4TCNQ & 16.5  & 9.5  & 4.4 & 9.0  \\ 
\hline
\end{tabular}
\end{table}

\FloatBarrier
\subsection{Appendix 2: Data Partitioning and Algorithm Performance Result}
\label{appendix2}
\FloatBarrier

\begin{table}[ht]
    \centering
    \caption{Kolmogorov--Smirnov (KS) tests comparing the empirical distributions of the training set with the validation and test sets for each parameter. The null hypothesis ($H_0$) is that the two samples are drawn from the same underlying distribution. For all parameters, the p-values exceed 0.05; therefore, $H_0$ is not rejected, indicating no statistically significant distributional shift between the splits.}
    \label{tab:ks}
    \begin{tabular}{lccccl}
        \hline
        \multirow{2}{*}{Parameter} & \multicolumn{2}{c}{KS Statistic} & \multicolumn{2}{c}{p-value} & \multirow{2}{*}{Comment} \\
        \cline{2-5}
        & Val & Test & Val & Test & \\
        \hline
        \% CB                      & 0.23 & 0.18 & 0.48 & 0.78 & Fail to reject $H_0$ \\
       
        \% DCB                     & 0.21 & 0.23 & 0.56 & 0.53 & Fail to reject $H_0$ \\
        
        \% Tol                     & 0.19 & 0.31 & 0.73 & 0.18 & Fail to reject $H_0$ \\
        
        Annealing Temp ($^\circ$C) & 0.15 & 0.21 & 0.92 & 0.61 & Fail to reject $H_0$ \\
        
        Conductivity (S/cm)        & 0.24 & 0.17 & 0.44 & 0.86 & Fail to reject $H_0$ \\
        \hline
    \end{tabular}
\end{table}

\begin{table}[ht]
    \centering
    \caption{{QSPR models' performance metrics for test dataset. 8 different machine learning algorithms were tried. Tree-based machine learning algorithms worked better than other classes of machine learning algorithms}}
    \label{tab:all_model_results}  

    \resizebox{\textwidth}{!}{%
    \begin{tabular}{ccccccccccc}
        \Xhline{1.2pt}
        \multirow{2}{*}{Type} & \multirow{2}{*}{Model} & \multirow{2}{*}{Algorithm} & \multirow{2}{*}{Input} & \multirow{2}{*}{Output} & $R^2$  & RMSE & MAE & Kendall  & Pearson & \multirow{2}{*}{Comment} \\
        & & &  & & (\% $\uparrow$) & ($\downarrow$)& ($\downarrow$) & Tau (\% $\uparrow$) & (\% $\uparrow$) \\
        \Xhline{1.2pt}
        \multirow{16}{*}{\shortstack{Data\\Driven}} 
        & \multirow{8}{*}{I-QSPR 1} & RF & \multirow{8}{*}{AUC,$p$} & \multirow{8}{*}{$\sigma$} 
            & \textbf{73.17} & \textbf{6.25} & 4.56 & 78.79 & \textbf{88.20} & Selected \\
            & & GB & & & 59.05 & 7.72 & 4.83 & 75.76 & 77.67 & \\
            & & Linear & & & 51.45 & 8.41 & 5.93 & 66.67 & 71.75  &\\
            & & LASSO & & & 57.45 & 7.87 & 5.63 & 69.70 & 77.03  &\\
            & & KR & & & 21.57 & 10.68 & 6.80 & 63.64 & 64.01  & \\
            & & SVR & & & 64.84 & 7.15 & \textbf{4.27} & 78.79 & 84.35  & \\
            & & kNN & & & 50.30 & 8.50 & 5.43 & \textbf{81.82} & 72.69  & \\
            & & GPR & & & 26.64 & 10.33 & 7.12 & 57.58 & 60.90  & \\

        \Xcline{2-11}{0.8pt}
        & \multirow{8}{*}{I-QSPR 2} & RF & \multirow{8}{*}{AUC,$p$, $M$} & \multirow{8}{*}{$\sigma$} &
            \textbf{73.18} & \textbf{6.25} & \textbf{4.39} & 75.76 & \textbf{88.74}  & Selected\\
            & & GB & & & 63.93 & 7.24 & 4.63 & \textbf{78.79} & 81.50  & \\
            & & Linear & & & 28.61 & 10.19 & 7.66 & 51.52 & 60.56  & \\
            & & LASSO & & & 66.73 & 6.96 & 5.08 & 66.67 & 82.43  &\\
            & & KR & & & -45.97 & 14.57 & 10.57 & 60.61 & 54.85  & \\
            & & SVR & & & 62.96 & 7.34 & 4.72 & 72.73 & 80.67  & \\
            & & kNN & & & 41.71 & 9.21 & 6.13 & 57.58 & 64.88  & \\
            & & GPR & & & 29.43 & 10.13 & 6.68 & 54.55 & 64.69  & \\

        \Xhline{1 pt}
        \multirow{8}{*}{Expert} 
        & \multirow{8}{*}{E-QSPR } & RF & \multirow{8}{*}{$E$} & \multirow{8}{*}{$\sigma$} 
            & 72.90 & 6.28 & 4.08 & \textbf{84.85} & 93.60 & \\
            
            & & GB & & & \textbf{81.49} & \textbf{5.19} & \textbf{3.49} & \textbf{84.85} & \textbf{94.53}  & Selected\\
            & & Linear & & & 36.49 & 9.61 & 6.18 & 66.67 & 63.19  & \\
            & & LASSO & & & 40.94 & 9.27 & 5.14 & 57.58 & 67.10  & \\
            & & KR & & & 25.65 & 10.40 & 7.03 & 63.64 & 72.24  & \\
            & & SVR & & & 55.82 & 8.02 & 4.50 & 72.73 & 83.66  & \\
            & & kNN & & & 31.56 & 9.98 & 6.57 & 60.61 & 56.97  & \\
            & & GPR & & & 31.89 & 9.96 & 6.71 & 54.55 & 63.74  & \\

        \Xhline{1.2pt}
    \end{tabular}%
    }

    \vspace{4pt}
    \caption*{
    \small
    \textbf{Details:} I-QSPR 1, I-QSPR 2: Intermediate models using data-driven features. 
    E-QSPR: Expert-curated model. 
    \\
    AUC: area-under-the-curve features from spectra and their second derivative; 
    $p$: processing conditions; 
    $\sigma$: conductivity; 
    $M$: interaction products between AUC features; 
    $D$: SHAP-selected data-driven subset of AUC, $p$, and $M$; 
    $E$: expert-identified features; 
    $C$: SHAP-selected best subset from $D$ and $E$.
    \\
    RF: Random Forest\\
    GB: Gradient Boosting\\
    KR: Kernel Ridge Regression\\
    SVR: Support Vector Regression\\
    kNN: k-Nearest Neighbor Regression\\
    GPR: Gaussian Process Regression

    We present results for I-QSPR 1, I-QSPR 2, and E-QSPR, as these models represent different stages of feature development and provide a valuable basis for comparison. I-QSPR 3 builds directly on I-QSPR 2. The best algorithm from I-QSPR 2 is chosen, and then we perform SHAP-based feature ranking and selection. A similar method is employed for the final QSPR model, which combines data-driven and expert-curated features and utilizes SHAP-based feature selection again. The comparison shows that tree-based models consistently outperform other model types across all evaluation metrics. Therefore, we based the next models (I-QSPR~3 and the final QSPR) on refining the tree-based approach. 
    }

\end{table}

\textbf{Hyperparameters}

\textbf{I-QSPR 1}: \texttt{n\_estimators} = 70, \texttt{criterion} = \texttt{squared\_error}, \texttt{min\_samples\_split} = 5

\textbf{I-QSPR 2}: \texttt{n\_estimators} = 50, \texttt{criterion} = \texttt{squared\_error}, \texttt{min\_samples\_split} = 2

\textbf{I-QSPR 3}: \texttt{n\_estimators} = 50, \texttt{criterion} = \texttt{squared\_error}, \texttt{min\_samples\_split} = 2

\textbf{E-QSPR}: \texttt{loss} = \texttt{squared\_error}, \texttt{learning\_rate} = 0.1, \texttt{n\_estimators} = 100, \texttt{min\_samples\_leaf} = 1

\textbf{QSPR}: \texttt{loss} = \texttt{squared\_error}, \texttt{learning\_rate} = 0.1, \texttt{n\_estimators} = 150, \texttt{min\_samples\_leaf} = 5

\FloatBarrier
\subsection{Appendix 3: Model Performance under Spectral Noise}
\label{appendix3}
\FloatBarrier

We added random 10\% Gaussian noise to the spectral data. We use the genetic algorithm-based optimization to obtain the bin locations. The bin locations obtained were [1.966, 1.76,  2.16, 2.51, 2.88] eV. The bin locations obtained from the original data were [1.378, 1.828, 1.982, 2.095, 2.700] eV. We observe that, barring the first location of 1.378 vs 1.966, the bins more or less cover the same spectral area. We use the bin location obtained from the noisy data and train our data-driven models. We observe that the model performance of the models based on original data was between 73 -76\%, and for the models based on bin location obtained from noisy data, it was between 74 -77\%, as shown in Table \ref{tab:modelnoisy_results}.

\begin{table}[ht]
    \centering
    \caption{{QSPR Models' Performance Metrics for Orignal and 10\% Noisy data}}
    \label{tab:modelnoisy_results}  

    \resizebox{\textwidth}{!}{%
    \begin{tabular}{cccccccccc}
        \Xhline{1.2pt}
        \multirow{2}{*}{Model} & \multirow{2}{*}{Type} & \multirow{2}{*}{Algorithm} & \multirow{2}{*}{Input} & \multirow{2}{*}{Output} & $R^2$  & RMSE & MAE & Kendall  & Pearson  \\
        & & & & & (\% $\uparrow$) & ($\downarrow$)& ($\downarrow$) & Tau (\% $\uparrow$) & (\% $\uparrow$) \\
        \Xhline{1.2pt}
         \multirow{2}{*}{I-QSPR 1} & Orignal & \multirow{2}{*}{\makecell{Random\\Forest}} & \multirow{2}{*}{AUC,$p$} & \multirow{2}{*}{$\sigma$} 
            & 73.17 & 6.25 & 4.56 & 78.79 & 88.20\\
     & 10\% Noise & & & & 77.26 & 5.75 & 4.22 & 78.79 & 91.81 \\

        \Xcline{1-10}{0.8pt}
        \multirow{2}{*}{I-QSPR 2} & Original & \multirow{2}{*}{\makecell{Random\\Forest}} & \multirow{2}{*}{AUC,$p$, $M$} & \multirow{2}{*}{$\sigma$} 
            & 73.18 & 6.25 & 4.39 & 75.76 & 88.74 \\
            & 10\% Noise & & & & 74.80 & 6.06 & 4.13 & 78.79 & 90.49 \\
         
        \Xcline{1-10}{0.8pt}
         \multirow{2}{*}{I-QSPR 3} & Orignal & \multirow{2}{*}{\makecell{Random\\Forest}} & \multirow{2}{*}{$D$} & \multirow{2}{*}{$\sigma$} 
            & 76.09 & 5.90 & 4.42 & 78.79 & 89.52 \\

         & 10\% Noise & & & & 77.87 & 5.67 & 4.01 & 72.73 & 91.26\\

        \Xhline{1.2pt}
    \end{tabular}%
    }

    \vspace{4pt}
    \caption*{
    \small
    \textbf{Details:} I-QSPR 1, I-QSPR 2, I-QSPR 3: Intermediate models using data-driven features. \\
    AUC: area-under-the-curve features from spectra and their second derivative; 
    $p$: processing conditions; 
    $\sigma$: conductivity; 
    $M$: interaction products between AUC features; 
    $D$: SHAP-selected data-driven subset of AUC, $p$, and $M$; 
    $E$: expert-identified features; 
    $C$: SHAP-selected best subset from $D$ and $E$.}

\end{table}

\FloatBarrier

\subsection{Appendix 4: Model Performance on Data with Conductivity over 30 S/cm}
\label{appendix 4}
\FloatBarrier
\begin{table}[ht]
    \centering
    \caption{{QSPR Models' Prediction for Conductivity Data over 30 S/cm in Validation and Test Set}}
    \label{tab:modelsbove 30_results}  

    \resizebox{\textwidth}{!}{%
    \begin{tabular}{ccccc}
        \Xhline{1.2pt}
        \multirow{2}{*}{Data} & True Conductivity & I-QSPR 1 Pred & I-QSPR 2 Pred & E-QSPR Pred\\
        & S/cm & S/cm & S/cm & S/cm\\
        \Xhline{1.2pt}

Val & 32.42 & 24.17 & 25.09 & 22.10\\ 
Val & 31.29 & 23.59 & 22.42 & 23.44\\
Val & 32.65 & 25.44 & 25.85 & 25.13\\
Val & 30.93 & 22.95 & 23.51 & 29.24\\
Test & 49.87 & 33.48 & 32.19 & 34.35\\
\Xhline{0.8pt}
 & MAE & 9.51 & 9.62 & 8.58 \\
 & MAE without Sample 4 & 9.88 & 10.17 & 10.30 \\
     
        \Xhline{1.2pt}
    \end{tabular}%
    }

    \vspace{4pt}
    \caption*{
    \small
    \textbf{Details:} I-QSPR 1, I-QSPR 2: Intermediate models using data-driven features. E-QSPR: Expert curated model } 
    
\end{table}
\FloatBarrier

\subsection{Appendix 5: SHAP Results for I-QSPR 3}
\label{appendix5}
\FloatBarrier

\begin{figure}[htbp]
	\centering
\includegraphics[width=1\linewidth]{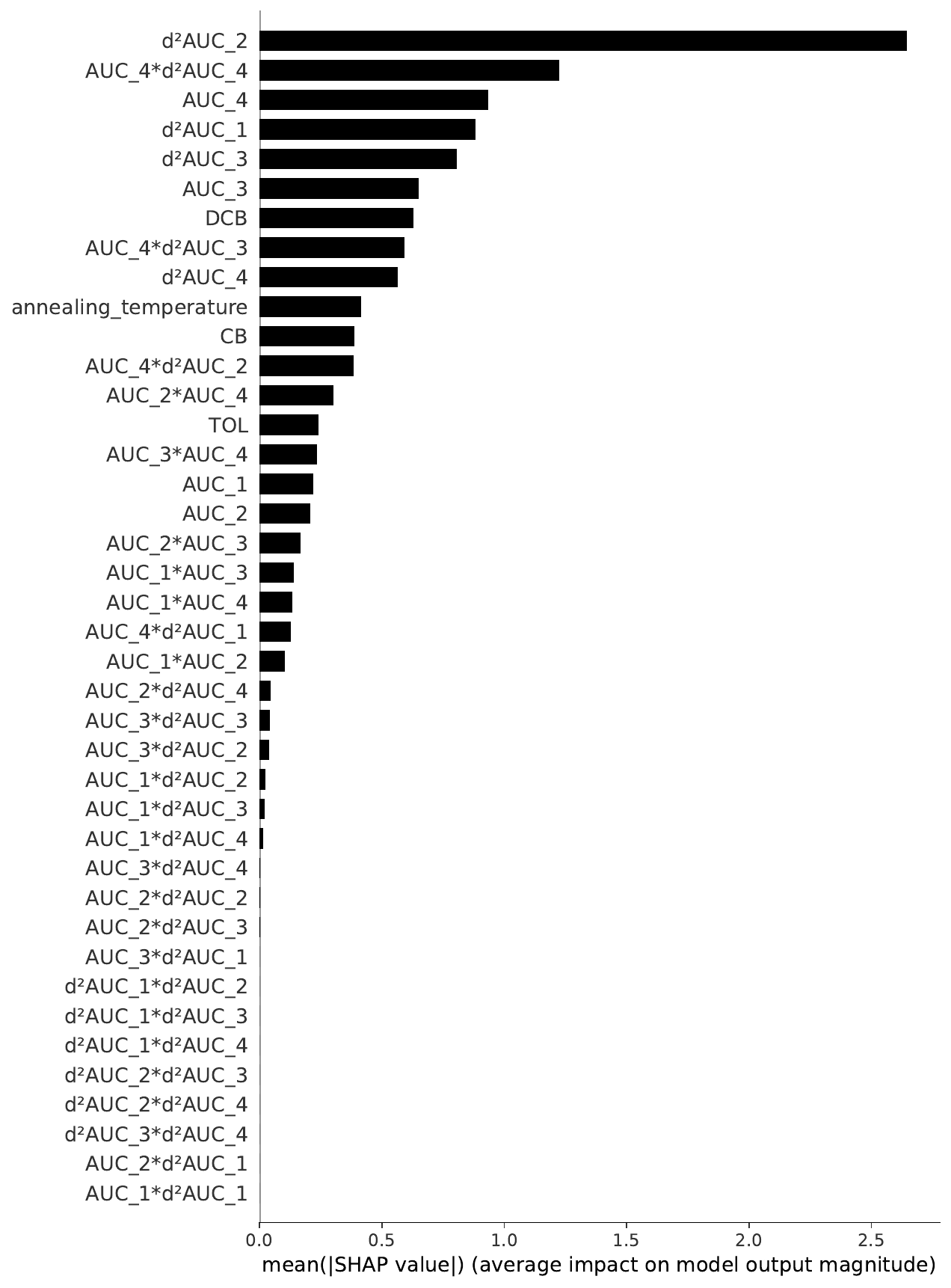}
    \caption{Feature importance (SHAP score) for each feature in I-QSPR 2}
    \label{Fig:shap_bar_all}
\end{figure}

\FloatBarrier

\subsection{Appendix 6: Expert Feature Terminology}
\label{appendix6}
\FloatBarrier

\begin{figure*}[t!]
	\centering
\includegraphics[width=1\linewidth]{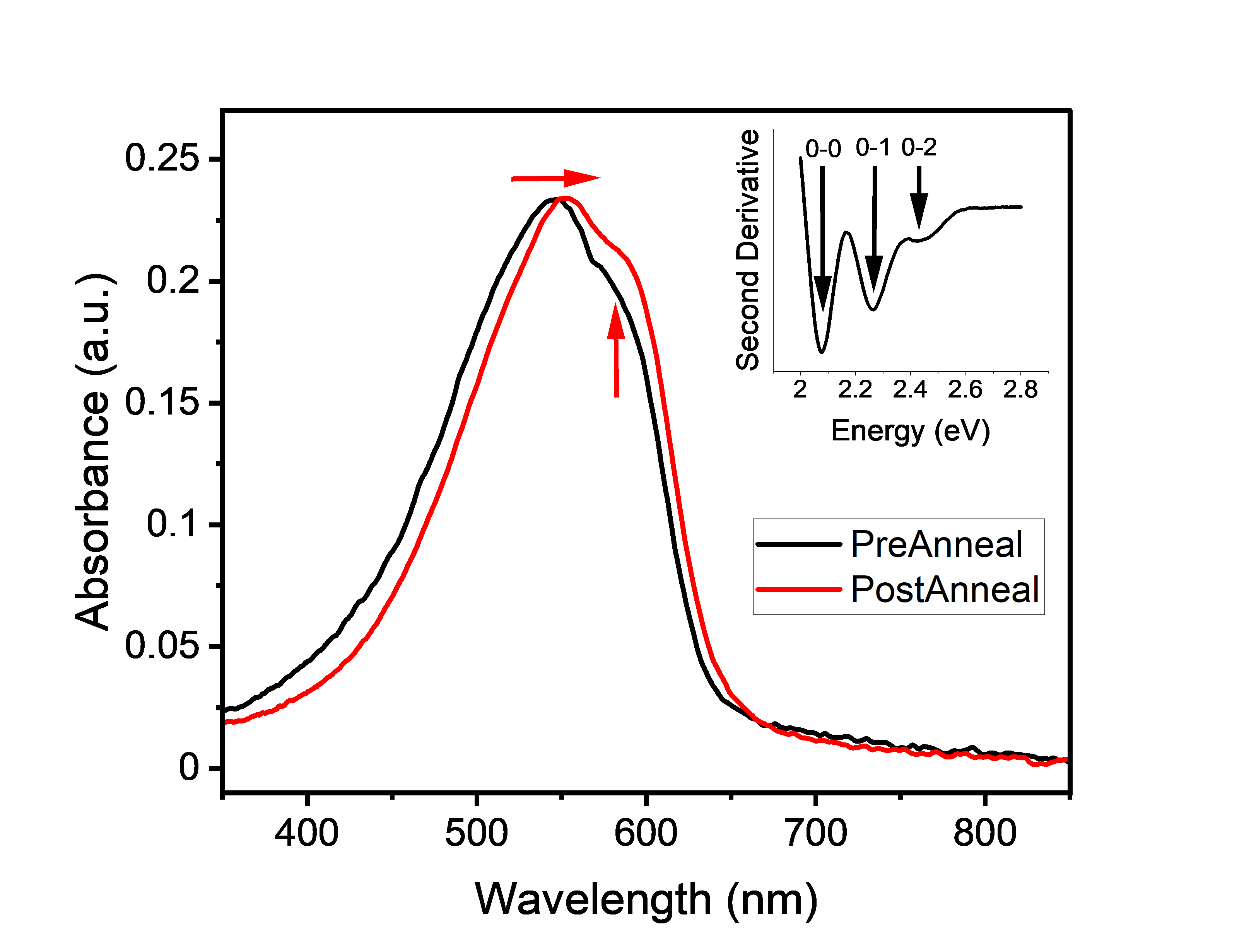}
    \caption{Example absorbance spectrum from a pBTTT film before and after annealing. Notable differences in the peak shifting and intensity highlight the effect of annealing and demonstrate some of the traditional features studied. The inset shows the second derivative of the absorption spectrum, which is used to identify the location of the 0-0, 0-1, and 0-2 vibronic transitions.}
    \label{Fig:vibronics}
\end{figure*}
Aggregation: The process by which individual polymer chains physically come together, often through \(\pi\)-\(\pi\) stacking or van der Waals forces. Aggregation can lead to changes in optical properties, such as red-shifted absorption or emission, due to increased interactions between chains. Differences in aggregation arising from co-solvent and/or annealing are often reflected in the absorption spectroscopy as noted in Figure \ref{Fig:vibronics}.

Red-shift: A shift of an absorption or emission peak to longer wavelengths (lower energy). Often indicative of stronger intermolecular interactions, increased conjugation length, or higher degrees of aggregation or planarity. Figure \ref{Fig:vibronics} shows a red shifting resulting from annealing.

Blue-shift: A shift of an absorption or emission peak to shorter wavelengths (higher energy). Often resulting from decreased conjugation length, structural disorder, disruption of aggregation, or increased localization of the excited state.

Vibronic Transition: An electronic transition that occurs along with a change in the molecule’s vibrational state. Common vibronic transitions are labeled 0-0, 0-1, and 0-2, where the first number refers to the vibrational level in the ground state and the second refers to the vibrational level of the excited state. Figure \ref{Fig:vibronics} inset shows how these transitions are found using the local minima in the second derivative of the absorption spectrum.

0-0 Transition: A transition between the lowest vibrational level of the ground state and the lowest vibrational level of the excited state. It represents pure electronic excitation and is often the most direct indicator of the intrinsic energy gap in a conjugated polymer.

0-1 Transition: A transition from the ground vibrational level of the ground electronic state to the first vibrational level of the excited electronic state.

0-2 Transition: A transition from the ground vibrational level of the ground electronic state to the second vibrational level of the excited electronic state.

Structural Order / Disorder: Refers to the degree of regularity or conformational alignment within a polymer assembly. Structural order tends to enhance electronic delocalization and sharpens optical features. Disorder often introduces broadening and increased vibronic progression.

Planarity: Refers to how flat or co-planar the backbone of a conjugated polymer is. Higher planarity facilitates better \(\pi\)-conjugation and delocalization, leading to sharper spectral features and improved charge transport. Planarity is a factor of structural order/disorder.

Delocalization: The extent to which an electronic excitation (e.g., exciton) spreads over multiple molecular units or chains. Delocalized excitons typically result in higher 0-0 transition prominence and narrower peaks, while localized excitons show stronger 0-1 and 0-2 vibronic progression.

Electron–Vibrational Coupling (Electron–Phonon Coupling): The interaction between an electron’s movement and vibrations of the molecule. Strong coupling leads to vibronic progressions (e.g., prominent 0-1, 0-2 peaks) and structural relaxation in excited states.

Vibronic Progression: The pattern of multiple vibronic peaks (e.g., 0-0, 0-1, 0-2…) in a spectrum that reflects the strength of vibrational coupling. A pronounced progression suggests stronger electron–vibration interactions.

Huang–Rhys Factor (S): A dimensionless quantity that quantifies electron–phonon coupling of a material.
A small S indicates weak coupling, often reflected in a sharp 0-0 peak, whereas a large S arises from strong coupling and is observed by more intense 0-1/0-2 transitions.
\FloatBarrier

\subsection{Appendix 7: Correlation between Data-Driven and Expert Features}
\label{appendix7}
\FloatBarrier
\begin{figure}[H]
    \centering
    \includegraphics[width=1\linewidth]{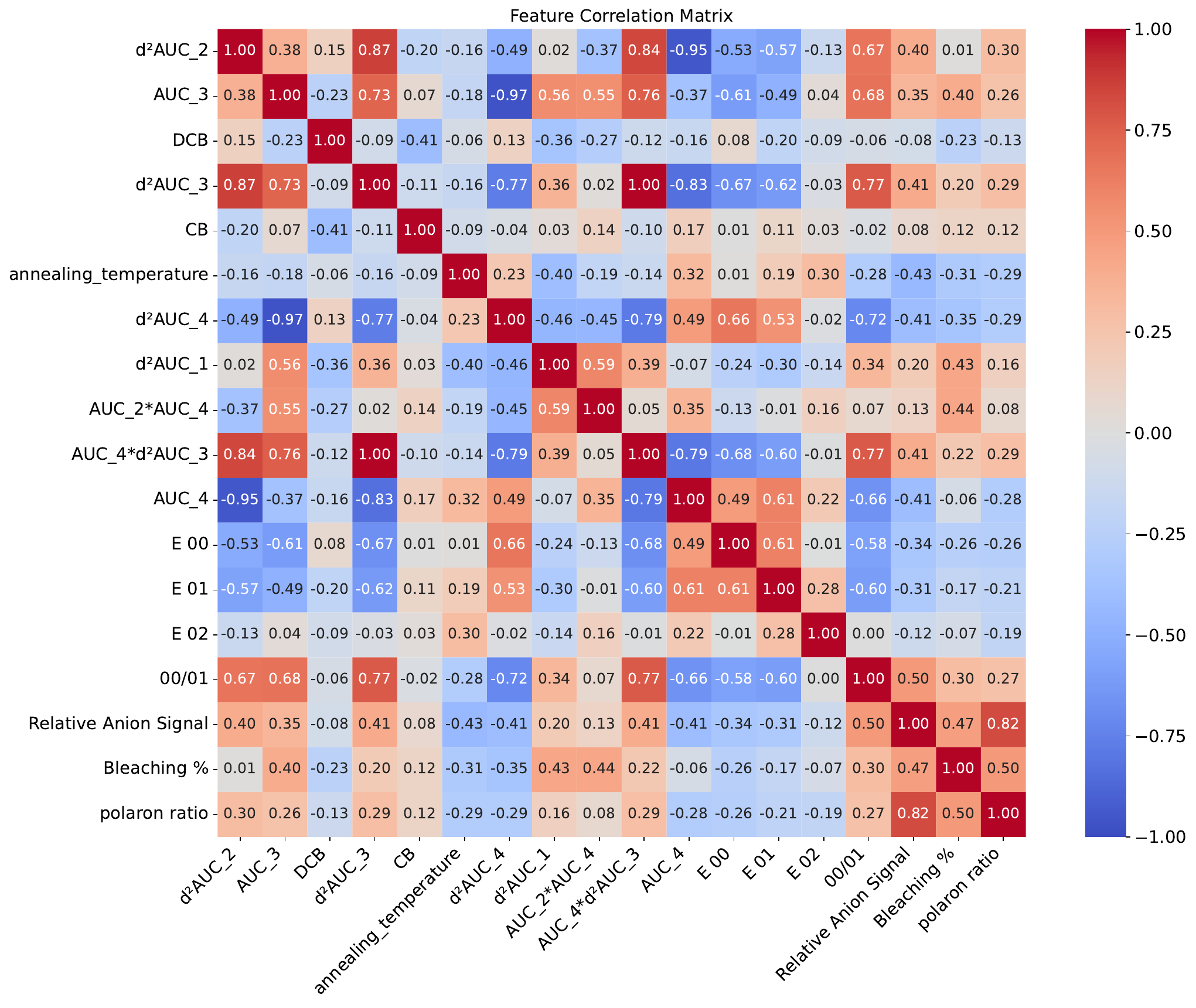}
    \caption{Spearman correlation between data-driven features (first 11 features) and expert-identified features (last 7 features)}
    \label{Fig:correlation_all}
\end{figure}
\FloatBarrier

\end{document}